\documentclass[reprint,superscriptaddress,nofootinbib,amsmath,amssymb,aps]{revtex4-1}

\usepackage{graphicx}
\usepackage{dcolumn}
\usepackage{bm}

\usepackage{cancel}

\newcommand{\Mstar}{\ensuremath{m_{\ast}}}
\newcommand{\Nstar}{\ensuremath{N_{\ast}}}

\begin{document}

\preprint{APS/123-QED}

\title{X-MRIs: Extremely Large Mass-Ratio Inspirals}

\author{Pau Amaro-Seoane}
\affiliation{Institute of Space Sciences (ICE, CSIC) \& Institut d'Estudis Espacials de Catalunya (IEEC) at Campus UAB, Carrer de Can Magrans s/n 08193 Barcelona, Spain}
\affiliation{Kavli Institute for Astronomy and Astrophysics at Peking University, 100871 Beijing, China}
\affiliation{Institute of Applied Mathematics, Academy of Mathematics and Systems Science, Chinese Academy of Sciences, Beijing 100190, China}
\affiliation{Zentrum f{\"u}r Astronomie und Astrophysik, TU Berlin, Hardenbergstra{\ss}e 36, 10623 Berlin, Germany}

\date{\today \\\vspace{0.2cm} \centerline{\textbf{For my dear friend Tal Alexander. Thanks for having been a human being.}}\vspace{0.05cm}}

\begin{abstract}
The detection of the gravitational waves (GWs) emitted in the capture process
of a compact object by a massive black hole (MBH) is known as an extreme-mass ratio
inspiral (EMRI) and represents a unique probe of gravity in the strong regime and
is one of the main targets of the Laser Interferometer Space Antenna (LISA).
The possibility of observing a compact-object EMRI at the
Galactic Centre (GC) when LISA is taking data is very low.
However, the capture of a brown dwarf (BD), an X-MRI, is more frequent because these
objects are much more abundant and can plunge without being tidally disrupted.
An X-MRI covers some $\sim 10^8$ cycles before merger, and hence stay on band
for millions of years.  About $2\times 10^6$ yrs before merger they have a
signal-to-noise ratio (SNR) at the GC of 10. Later, $10^4$ yrs before merger, the SNR is of several thousands, and
$10^3$ yrs before the merger a few $10^4$. Based on these values, this kind of
EMRIs are also detectable at neighbour MBHs, albeit with fainter SNRs. We
calculate the event rate of X-MRIs at the GC taking into account the
asymmetry of pro- and retrograde orbits on the location of the last stable
orbit.
We estimate that at any given moment, and using a conservative approach, there
are of the order of $\gtrsim\,20$ sources in band. From these, $\gtrsim\,5$ are
highly eccentric and are located at higher frequencies, and about $\gtrsim\,15$
are circular and are at lower frequencies.
Due to their proximity, X-MRIs represent a unique probe of gravity in the
strong regime.  The mass ratio for a X-MRI at the GC is $q \sim 10^8$, i.e.,
three orders of magnitude larger than stellar-mass black hole EMRIs. Since
backreaction depends on $q$, the orbit follows closer a standard geodesic,
which means that approximations work better in the calculation of the orbit.
X-MRIs can be sufficiently loud so as to track the systematic growth of their
SNR, which can be high enough to bury that of MBH binaries.

\end{abstract}

\maketitle

\section{Introduction}
\label{sec.intro}

Thanks to major advances in high angular instrumentation we have observed a
fundamental link between the features of a host galaxy and those its central
MBH \citep{KormendyHo2013}. The lower end of the so-called mass-sigma
correlation remains uncertain, but if we assume that it remains valid, then
smaller dense stellar systems, such as globular clusters should also contain
black holes, although in a smaller mass range. These are known as
intermediate-mass black holes, IMBHs, and are supported by the existence of
ultra-luminous X-ray emission \cite{Mezcua2017,LuetzgendorfEtAl2013}.

An excellent probe of I/MBHs are the GWs emitted by the slow inspiral of a
sufficiently compact stellar object which radiates energy away and
slowly approaches the I/MBH. This is called an extreme- or
intermediate-mass ratio inspiral, depending on the mass-ratio between the compact object and
the I/MBH (EMRIs, $\gtrsim 10^4:1$ or IMRIs, $\sim 10^2-10^4:1$).
EMRIs and IMRIs can be detected by the Laser Interferometer Space Antenna
(LISA) mission
\citep{Amaro-SeoaneEtAl2017,Amaro-SeoaneLRR2012,Amaro-SeoaneGairPoundHughesSopuerta2015}
as well as by ground-based detectors such as the Laser Interferometer
Gravitational-Wave Observatory, in principle jointly with LISA
\citep{Amaro-Seoane2018}.

Compact objects can plunge through the event horizon under the assumption that the stellar object can
withstand the enormous tidal forces exerted on it.  Indeed, if the object is
an extended star such as our Sun, some or all of it (depending on
the distance of minimum approach) may be torn apart because of the tidal
gravity of the central object \citep{Hills75,Rees88}. The difference in
the gravitational force on points diametrically separated on the star alter its
shape, from its initial approximately spherical architecture to an ellipsoidal
one and, in the end, the star is disrupted. This occurs whenever
the work exerted over it by the tidal force exceeds its own binding energy.

\begin{figure}
\resizebox{\hsize}{!}
          {\includegraphics[scale=1,clip]{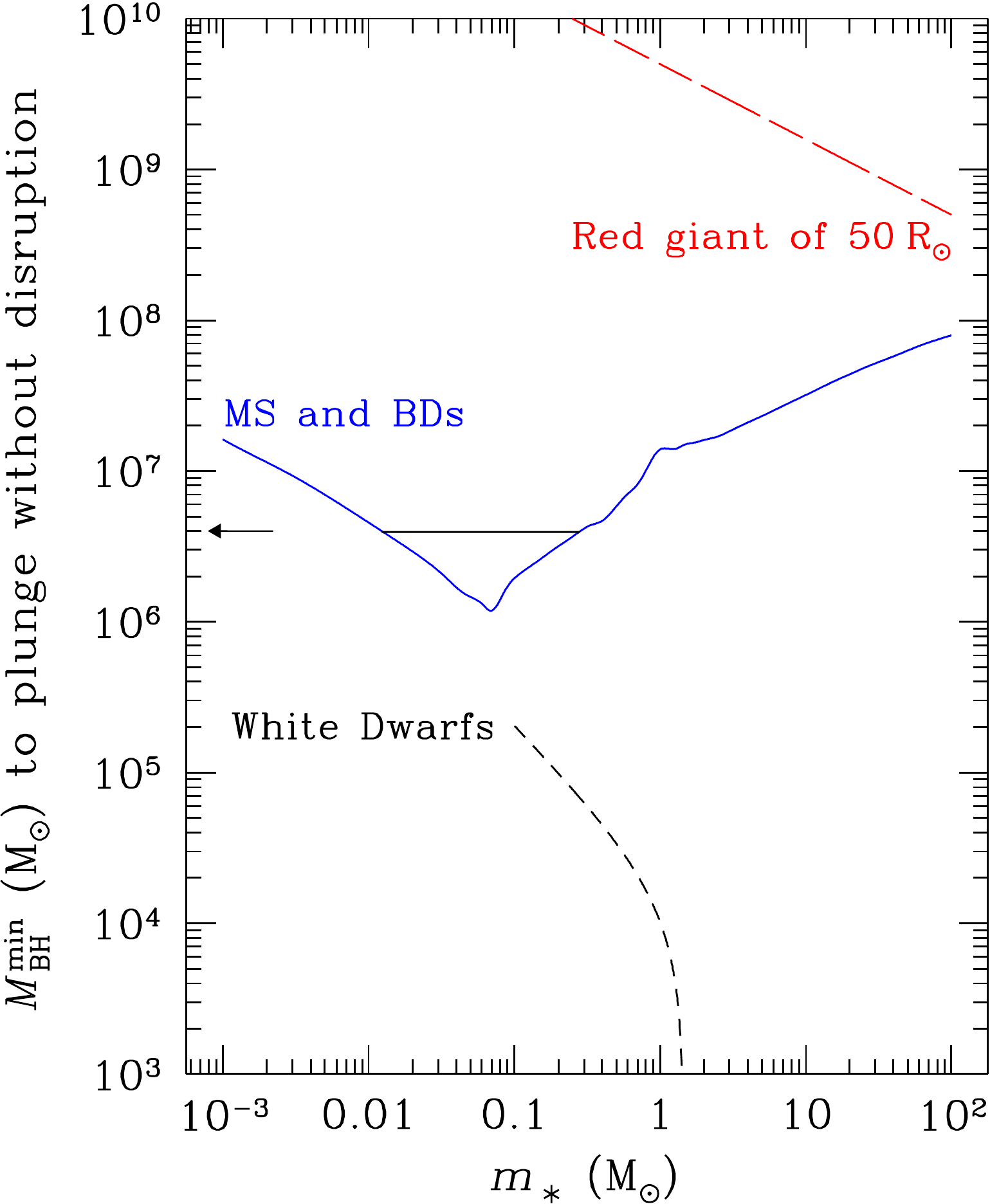}}
\caption
   {
Minimum mass for a MBH for a given stellar object to plunge through the event horizon without a tidal disruption as a function of its mass $m_{\ast}$.
For each different kind of stellar object --- i.e., red giants, main sequence stars, and sub-stellar objects, we use realistic mass-ratio relations (see text).
SgrA*, with a mass of about $\sim 4\times 10^6\,M_{\odot}$, is marked with an
arrow, which indicates the range of masses of sub-stellar objects that
might cross the event horizon without significant tidal stresses.
   }
\label{fig.M_R}
\end{figure}

Whether or not a stellar object can successfully cross the event horizon of the MBH
without being tidally disrupted can be estimated by equating the plunge
and the tidal radii. We adopt the approximation in Newtonian mechanics of
\cite{ST83} based on the estimation of a critical angular momentum $J_{\rm
crit}<4GM_{\rm BH}c^{-1}$ that leads to a successful inspiral of a particle at
infinity. The authors show that $J_{\rm crit}$ defines a parabolic orbit or
pericentre distance

\begin{equation}
R_{\rm p}:=4\,R_{\rm S}= 8\frac{GM_{\rm BH}}{c^2},
\label{eq.Rp}
\end{equation}

\noindent
which we adopt as the plunge radius. In this equation
$G$ is the gravitational constant and $c$ the speed of light in vacuum.
The tidal radius of the stellar object can be defined as

\begin{equation}
r_{\rm t} =\left(2\,\frac{5-n}{3\,m_{\ast}} {M}_{\rm BH}\right)^{1/3} r_{\ast},
\label{eq.r_tid_bind}
\end{equation}

\noindent
where $r_{\ast}$ and $m_{\ast}$ are the radius and mass of the star,
respectively, and  $n$ the polytropic index \cite{Chandra42}.  Hence, by
equating Eqs.~(\ref{eq.Rp}) and (\ref{eq.r_tid_bind}), we can obtain a
threshold mass for the MBH above which stellar objects plunge through the event
horizon without suffering significant tidal stresses on their structure,
$M_{\rm BH}^{\rm min}$. In Fig.~(\ref{fig.M_R}) we show this mass as a function
of the mass of various kinds of stellar objects, ranging from red giants to
main-sequence and objects such as brown dwarfs, to white dwarfs. What
determines a successful plunge is the mass-radius relation of the given object.
In Fig.~(\ref{fig.M_R}), we use the $m_\ast$--$r_\ast$ relations from detailed
modelling of stars given in \citep{SSMM92,MMSSC94,CDSBMMM99,CB00}, and note
that the relations provided in those articles for sub-stellar objects reproduce
almost exactly the more recent results of \cite{ChabrierEtAl2009} for
brown-dwarfs.  In Fig.~(\ref{fig.M_R}) we see that these ojects can plunge
through the event horizon of MBHs of masses similar to the one in our Milky Way
without being tidally torn apart. These objects are potential sources of
inspirals with an extremely large mass ratio, of $\sim\,10^8$, which translates
into a very large number of cycles before merger. Due to their proximity,
the signal-to-noise ratios can be as large as $20,000$. Because of these extreme
properties, in this paper we call BD EMRIs ``X-MRIs''\footnote{As suggested
by Bernard Schutz to us last X-mas.}.

The possibility of having bursts of gravitational radiation sources in our GC
from main-sequence (i.e. extended) stars (which are eventually tidally
disrupted) and the MBH has been addressed with Monte Carlo simulations
\citep{Freitag03}, and \citep{BarackCutler2004} discussed over the implications
in Section V.C and their Fig.~11. {They include a study of the SNR that these
sources have, and find similar results to our work. However, they do not derive
the event rate and the number of sources in band.} Also,
\cite{GourgoulhonEtAl2019} have also addressed the emission of radiation in our
GC from main-sequence stars and BDs. They consider circular orbits but their
study is complete in the Kerr metric, and include tidal effects, which are
important for the main-sequence stars. {As before, they do not address
the event rate and number of sources in band, at any given time, which we
derive in this work.}

In this paper we address the gravitational capture of sub-stellar objects at
the Galactic Centre, we calculate their signal-to-noise ratio and derive a
merger event rate. Thanks to the rates, and taking into account the key
property of these sources, their mass ratio, we derive the number of sources in
the band of the space-borne LISA observatory as a function of their
eccentricity and signal-to-mass ratio, and discuss the implications for
observations.

\section{Distribution around SgrA*}

The quasi-steady solution for the distribution of stars around a MBH takes the
form of an isotropic distribution function in energy space $f(E)\sim E^{\,p}$,
which translates into $\rho(r)\sim r^{-\gamma}$ in terms of the stellar density
$\rho$ (see \cite{Peebles1972} and \cite{BW76}, but also \cite{Gurevich64} for
a similar solution for the distribution of electrons around a positively
charged Coulomb centre). This solution has been confirmed a number of times
using semi-analytical and numerical approaches; see, e.g.,
\cite{SM78,MS79,MS80,ST85,FB01a,ASEtAl04,PretoMerrittSpurzem04}. In particular,
for our Galactic Centre, see the numerical work of \citep{BaumgardtEtAl2018},
which describes very well the observational data of
\citep{Gallego-CanoEtAl2018,SchoedelEtAl2018}.  Therefore, we assume a
power-law mass distribution for the stellar density,

\begin{equation}
 \rho(r) = \rho_0 \left( \frac{r}{r_0} \right)^{-\gamma},
\end{equation}

\noindent
with $\gamma$ the exponent value, and $\rho_0$ the stellar density at a
characteristic radius $r_0$ of normalization. Since we are interested in power-law cusps, as we will discuss later,
this can be chosen to be the influence radius, $r_0 \equiv R_{\rm h}$, and is only valid for radii smaller than this value.
The enclosed mass $M(r)$ at a certain radius $r$ can be estimated by solving
the integral

\begin{equation}
 M(r) = 4\,\pi \int^{r}_{0} \rho(r')\,r'^2 dr',
\end{equation}

\noindent
so that, with the proviso that $\gamma<3$,

\begin{equation}
 M(r)=\frac{4\,\pi}{3-\gamma}\,\rho_0\,r_0^3\left(\frac{r}{R_{\rm h}}\right)^{(3-\gamma)}.
\end{equation}

\noindent
Main sequence (MS) stars build a power-law distribution about the MBH, so that we can set
$r_0=R_{\rm h}$ for them, the influence radius of the MBH at the
Galactic Centre. Also, $M_{\rm BH}:=(4\,\pi)/(3-\gamma) \cdot \rho_0\,r_0^3$ is
the mass of the MBH, we have

\begin{equation}
 M(r)=M_{\rm BH}\left(\frac{r}{R_{\rm h}}\right)^{(3-\gamma)}
\end{equation}

Assuming that the mean stellar mass $\bar{m}_{\ast}$ at the radii of interest is independent
of the radius, the number of stars at a given radius $r$ is

\begin{equation}
 N_{\ast}(r)=\frac{M(r)}{\bar{m}_{\ast}}=\frac{M_{\rm BH}}{\bar{m}_{\ast}}
\left(\frac{r}{R_{\rm h}}\right)^{(3-\gamma)}
\label{eq.Ntot}
\end{equation}

Hence, the number of a given sub-population with a number fraction $f_{\rm
sub}$ of the stars at a radius $r$ can be calculated with

\begin{equation}
 N_{\rm sub}(r) = f_{\rm sub} \frac{M_{\rm BH}}{\bar{m}_{\ast}}
\left(\frac{r}{R_{\rm h}}\right)^{(3-\gamma)}.
\label{eq.Nsubpop}
\end{equation}

We must take into account that by doing so we are implicitly assuming that
light stars (or substellar objects, the main interest of this work) follow the
same distribution as MS stars. In this approach, both the MS stars and the BDs
are the light stellar component with their own power-law index, different to
the power-law index of stellar-mass black holes, which build a more
concentrated distribution around the MBH. A more realistic representation would
require more than two different exponents, but then it would be very difficult
to treat the problem analytically, as we do in this article. While until now we
have used $\gamma$ as a generic exponent, from now on, we will use it only for
the exponent of the stellar-mass black hole population (for historical reasons
this has been the convention), and $\beta$ for the light star population, i.e.
the BDs.

To derive $\bar{m}_{\ast}$ and $f_{\rm sub}$, we have to take into account that
BDs have masses ranging from approximately $0.01-0.07\,M_{\odot}$ (which is
actually a lower limit, since they can also have masses in the range
$0.07-0.15\,M_{\odot}$ through the BD formation process, see
\cite{KroupaEtAl2013}) and have their own initial mass function (IMF), which
is not well known-, but can be approximated by a single power law; see Eq.
(4.55) of \cite{KroupaEtAl2013}, which is consistent with observational data of
the inner galaxy \citep{WeggEtAl2017}. The IMF we consider is the usual Kroupa
broken power-law of the form $d\Nstar/d\Mstar \propto \Mstar^{-\alpha}$. We use
the mass intervals $[0.01,\,0.07,\,0.5,\,150]\,\times M_{\odot}$ with
$\alpha=0.3,\,1.3,\,2.3$, because the bulge may have had a top-heavy IMF; see
\citep{BartkoEtAl10} and \citep{BalleroEtAl2007} for some constraints on the
top-heaviness in the bulge, althought it remains unknown if the IMF below
$1\,M_{\odot}$ was different (Pavel Kroupa, personal communication).  We hence
introduce a discontinuity in this IMF to mimic the discontinuity between the
sub-stellar (BDs) population and the stellar IMF at $0.07\,M_{\odot}$ (Fig.
4-23 of \citep{KroupaEtAl2013}).  Assuming these values, we find that for the
BDs, $f_{\rm sub} \sim 0.21$, and that the average stellar mass is
$\bar{m}_{\ast} \sim 0.27\,M_{\odot}$. Taking into account that the influence
radius of our MBH, SgrA*, is $R_{\rm h} \sim 3\,{\rm pc}$
\citep{SchoedelEtAl2014,SchoedelEtAl2018}, and adopting the value $\beta=1.5$ \citep{BW77}, we obtain that

\begin{equation}
 N_{\rm BD}(r) \cong 6\times 10^{5} r_{\rm pc}^{1.5},
\label{eq.Nbd}
\end{equation}

\noindent
with $r_{\rm pc}$ the considered radius in pc. Therefore, at a distance of
$10^{-3}\,{\rm pc}$, there should be around $\sim 20$ BD.
This is the number of sources as a function of radius; however, the the
phenomenon we are interested on--- the formation, evolution and merger of an
X-MRI with the central MBH--- also represents a drain on these sources.  In
order to have a statistical picture of what we might expect at the Galactic
Centre once LISA starts to gather data, we need to address the (relativistic)
loss-cone problem of this scenario. This will also allow us to derive an event
rate.

\section{Loudness}

Assuming that one of these objects could indeed be in the relevant part of
phase space and become a source, in order to assess whether these sources are
interesting for LISA, we estimate in this section the signal-to-noise ratio
(SNR).  At a distance D given, an EMRI with a power emitted $\dot E$ and a rate of
change of frequency $\dot{f}$, its characteristic amplitude $h_{\rm c}$ can be
defined as $h_{\rm c} = \sqrt{(2\dot E/\dot f)}/(\pi D)$
\citep{FinnThorne2000}. If we assume a perfect signal processing, the sky- and
orientation-averaged SNR is given by \citep{FinnThorne2000}

\begin{equation}
\left(\frac{S}{N}\right)^2 = \frac{4}{\pi D^2} \int \frac{\dot{E}}{\dot{f} \, S_h^{SA}(f)} \frac{{\rm d}f}{f^2},
\end{equation}

\noindent
with $S_h^{SA}(f) \approx 5 S_h(f)$ the noise spectral density of the detector.
In the case of the EMRI problem, we need to sum the previous expression over
each mode to obtain the total SNR$^2$, since the signal has multiple frequency
components.  In this article we consider quadrupolar gravitational radiation,
and approach the orbit as a Keplerian ellipse with parameters that evolve
slowly due to the emission of GWs, as presented in the work of
\citep{Peters64}. Following this approximation, we decompose the amplitude in a
series of harmonics. For typical values relevant to this article, the n-th
harmonic at a distance $D$ is given by

\begin{align}
    h_n &= g(n,e) \frac{G^2\,M_{\rm BH} m_{\rm BD}}{D\,ac^4} \nonumber \\
        \nonumber &\simeq 10^{-18} g(n,e)
    \left(\frac{D}{8\,\mathrm{kpc}}\right)^{-1}
    \left(\frac{a}{10^{-3}\,\mathrm{pc}}\right)^{-1} \nonumber \\
    & \left(\frac{M_{\rm BH}}{4\times10^6\,M_{\odot}}\right)
    \left(\frac{m_{\rm BD}}{0.05\,M_{\odot}}\right). \nonumber
\label{eq.harm}
\end{align}

\noindent
In this equation, $g(n,\,e)$ is a function of the harmonic number $n$, and $e$
the eccentricity. Also, we note that the root mean square is considered to be averaged over
the two polarizations and all directions.  We hence consider the contribution
of the different harmonics (see Eq. 2.1 of \citep{FinnThorne2000} and Eq.
56 of \cite{BarackCutler2004} for more details),

\begin{equation}
\left(\frac{S}{N} \right)^2_n = \int^{f_n(\rm t_{fin})}_{f_n(\rm t_{ini})}
\left[\frac{h_{\rm c,\,n}(f_n)}{h_{\rm det}(f_n)} \right]^2
{\frac{1}{f_n}\,d\left(\ln(f_n) \right)}.
\end{equation}

\noindent
In this equation $f_n(t)$ is the (in principle redshifted, but irrelevant for the
GC) frequency of the nth harmonic at time $t$ (with $f_n=n \cdot f_{\rm orb}$,
$f_{\rm orb}$ being the orbital frequency{, and we note that  there are two differing
orbital frequencies for an eccentric body, radial and azimuthal. 
In \cite{BarackCutler2004} a compromise between the two is introduced.}) and $h_{\rm c,\,n}(f_n)$ the
characteristic amplitude of the nth harmonic when the frequency associated to
that component is $f_n$. Finally, $h_{\rm det}$ is the square root of the
sensitivity curve of LISA and we note that $d\left(\ln(f_n)
\right)/f_n$ is simply $df_n$.

\begin{figure*}
\resizebox{\hsize}{!}
          {\includegraphics[scale=1,clip]{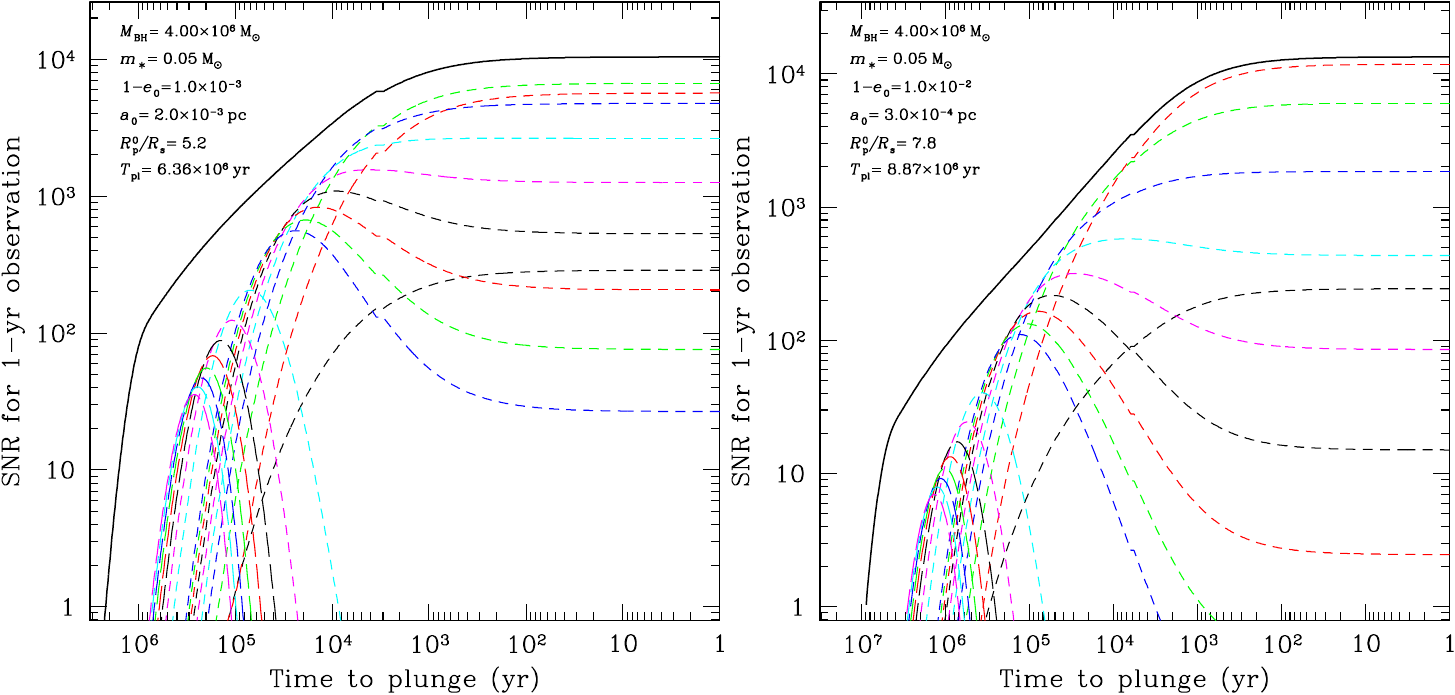}}
\caption
   {
SNR obtained by LISA for a one-year observation for an X-MRI of mass
$0.05\,M_{\odot}$ at the GC. At a given time, the curve shows the SNR we would
get if we followed the X-MRI for one year. The mass of the
MBH is set to $M_{\rm BH}=4\times10^{6}M_{\odot}$, while the initial eccentricity to
and semi-major axis are set to typical values. We show the initial pericentre distance in terms of the
Schwarzschild radius of the MBH and the (initial) time for the star to
plunge on to it under the assumption that it evolves only due to the loss of
energy in the form of GWs, as approximated by \citep{Peters64}.  We show the
contribution of the first 10 harmonics ({displayed in different colours, }and we note that we have used 1000 in
the calculation of the SNR). The black, solid line corresponds to the total
SNR and the red, dashed curve, to the second harmonic.
   }
\label{fig.BD_SNR}
\end{figure*}

\begin{figure*}
\resizebox{\hsize}{!}
          {\includegraphics[scale=1,clip]{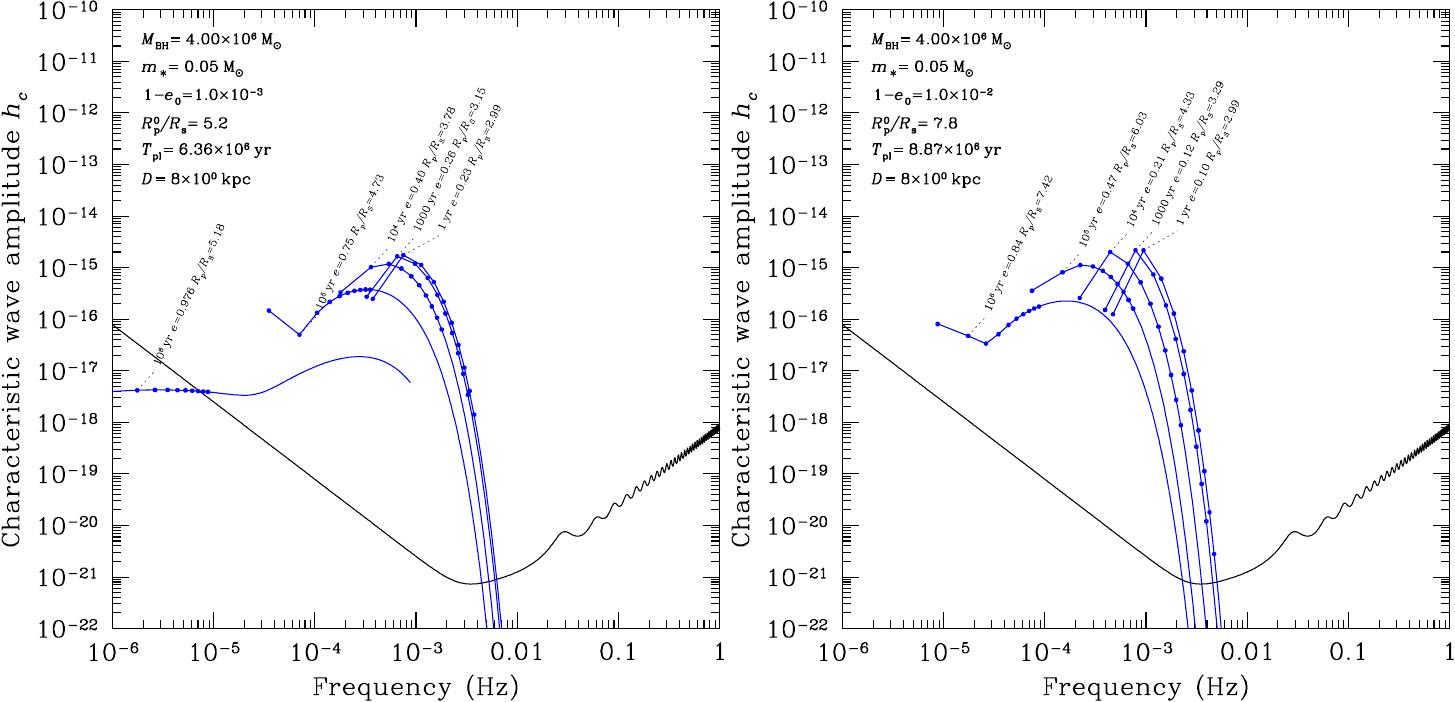}}
\caption
   {
Peak of the gravitational wave frequency emitted by the X-MRI systems of
Fig.~(\ref{fig.BD_SNR}).
Each blue line is an
isochrone made by selecting a given moment before the final plunge, shown as a
dashed label, for the first 1000 harmonics (see text), although we only depict the first ten ones {in each line, at the same time,} with circles.
{The curves are an interpolation from the 1000 harmonics.}
Additionally, we show the corresponding pericentre distance $R_{\rm p}$ in
units of the Schwarzchild radius $R_{\rm S}$ and the eccentricity. The time
$T_{\rm pl}$ is the time to merge as calculated from the initial dynamical
parameters of the binary, displayed at the top, left legend.
   }
\label{fig.isochr}
\end{figure*}

In Fig.~(\ref{fig.BD_SNR}) we give an example of an X-MRI of a given mass
for the BD object at the GC. The curves are to be interpreted
as the SNR that we would observe if we integrated for one year. Hence, at
a given time in the X axis, the SNR is what we would obtain if we followed the
source for one year-, i.e. if LISA could only operate for one year.  We can see
that from starting about $10^7~{\rm yrs}^{-1}$ before plunging through the event
horizon, these sources already have SNR $> 10$. If the sources were on their
last year of inspiral when LISA starts to get data, the SNR would be as much as
a few $10^3$ for the light BD in the left panel and $\sim 2\times 10^4$ for the
larger one in the right panel. We note that our calculation of the SNR is in
good agreement with the one done by \citep{BarackCutler2004}, which addressed
extended stars of low mass undergoing tidal disruptions at the GC, and the more
recent work of \citep{GourgoulhonEtAl2019}, which focuses on circular orbits.
In Fig.~(\ref{fig.isochr}) we show the peak of the frequency emitted by the same
systems.

\section{Event Rates}
\label{sec.Crossing}

The event rate, i.e., the number of X-MRIs that successfully inspiral and cross
the event horizon can be calculated by integrating in phase-space the number of
sources from a critical radius $a_{\rm crit}$ down to a minimum distance, $a_{\rm min}$,

\begin{equation}
\dot{\Gamma}_{\rm X-MRI} \simeq \int^{a_{\rm crit}}_{a_{\rm min}} \frac{dn_{\rm BD}(a)}
                     {T^{}_{\rm rlx}(a)\,\ln{\left(\theta_{\rm lc}^{-2}\right)}}\,.
\label{eq.EventRate}
\end{equation}

\noindent

We do not need to care about the specific shape of the curve in the integral,
because on the left of the LSO the integral will naturally vanish. To solve
this integral, we need (1) $\theta_{\rm ls}$, the loss-cone angle (see e.g.
\cite{Amaro-SeoaneLRR}), (2) $n^{}_{\rm BD}(a)$ the number of BDs within a
given semi-major axis $a$, (3) $T_{\rm rlx}(a)$ the relaxation time, and (4)
$a^{}_{\rm crit}$ and $a^{}_{\rm min}$ the critical and minimum radii, 
which is the upper and lower limits of the integral.

\subsection{The loss-cone angle}

The first quantity, (1) the loss-cone angle, can be estimated as \citep[][]{AL01}

\begin{equation}
 \theta_{\rm lc} \simeq \frac{1}{\sqrt{J^{}_{\rm max}/J^{}_{\rm lc}}}.
\label{eq.thetalc}
\end{equation}

\noindent
Using the same reference, we have that

\begin{equation}
J^{}_{\rm lc} \simeq \frac{4\,G}{c}\,M^{}_{\rm BH},\,J^{2}_{\rm max} = G M^{}_{\rm BH}a,
\end{equation}

\noindent
so that we obtain the first quantity to solve the integral,

\begin{equation}
\theta^{2}_{\rm lc} \simeq \sqrt{\frac{8\,R_{\rm S}}{a}}.
\label{eq.thetalc2}
\end{equation}

\subsection{Number of sources within a given radius}

Regarding the second quantity, (2) the number of BDs within a specific semi-major axis $a$,
we have already estimated in the previous section
how many BDs we might expect.
Hence, the number of BDs within $a$ is

\begin{equation}
N^{}_{\rm BD}(a) = f_{\rm sub}^{\rm BD}\cdot N^{\rm BD}_{0\,\textrm{MS}}\left(\frac{a}{R^{}_{0}}\right)^{3-\beta}\,.
\label{eq.Nbullet}
\end{equation}

\noindent
As discussed previously, $N^{\rm BD}_{0\,\textrm{MS}}$ is the total number of objects
(main-sequence stars and substellar objects) within $R_0$, which we choose to be $R_{\rm h}$, and $f_{\rm sub}^{\rm BD}$ is the number
fraction of BDs. In order to obtain the numerator in the integrand of
Eq.~(\ref{eq.EventRate}), we differentiate the last equation and obtain the
second quantity,

\begin{equation}
dn^{}_{\rm BD}(a) = f_{\rm sub}^{\rm BD}\,(3-\beta)\frac{N^{\rm BD}_{0\,\textrm{MS}}}{R^{}_{h}}
                   \left(\frac{a}{R_{h}}\right)^{2-\beta}da \,.
\end{equation}

\subsection{The relaxation time}

We can calculate the third quantity, (3) the relaxation time, by approximating
relaxation to be predominantly due to the population of stellar-mass black
holes of mass $m_{\rm bh}$, which dominate the central densities (e.g.
\cite{Amaro-SeoaneLRR}).

The relaxation time for a given distance which we take
to be equal to the semi-major axis $a$ is

\begin{equation}
T^{}_{\rm rlx}(a) = T^{}_{0}\left(\frac{a}{R^{}_{0}}\right)^{\gamma-3/2}\,,
\label{eq.Tr}
\end{equation}

\noindent
with \citep[][]{Spitzer87}

\begin{equation}
T^{}_{0} \sim 0.3389\, \frac{\sigma_{0}^3}{\ln(\Lambda)\,G^2\,m_{\rm bh}^2\,n^{}_{0}}\,,
\label{eq.t0}
\end{equation}

\noindent
and

\begin{align}
n^{}_{0}     & = \frac{3-\gamma}{4\pi}\frac{N^{}_0}{R_{0}^3} \label{eq.n0} \\
\sigma_{0}^2 & = \frac{1}{1+\gamma}\frac{G M^{}_{\rm BH}}{R^{}_{0}} \label{eq.sigma0} .
\end{align}

\noindent
In the previous equations $\ln(\Lambda)$ is the Coulomb logarithm, $G$ the gravitational constant,
$\sigma_0$ the velocity dispersion, $R_0$ is the radius within which stellar-mass black holes
dominate relaxation, and $N^{}_0$ the number of stellar-mass black holes enclosed in $R_0$ which,
in principle, can be smaller than the influence radius $R_{\rm h}$.
Hence, Eq.~(\ref{eq.t0}) becomes

\begin{equation}
T^{}_{0} \simeq \frac{4.26}{(3-\gamma)(1+\gamma)^{3/2}}
             \frac{\sqrt{R_{0}^3(G M^{}_{\rm BH})^{-1}}}{\ln (\Lambda)\,N^{}_{0}}
             \left(\frac{M^{}_{\rm BH}}{m^{}_{\rm bh}}\right)^2 \,.
\label{eq.t0full}
\end{equation}

Before we carry on with the rest of quantities necessary to solve
Eq.~(\ref{eq.EventRate}), we address two important points related to
Eq.~(\ref{eq.Tr}) and Eq.~(\ref{eq.t0full}).

First, (i) in order to derive $R_0$ of Eq.~(\ref{eq.Tr}), Eq.~(\ref{eq.n0}) and Eq.~(\ref{eq.sigma0}),
we note that the relaxation rate is (see e.g. \cite{Amaro-SeoaneLRR2012})

\begin{equation}
\Gamma_{\rm rlx} = \frac{32}{\pi\,v_{\rm rel}^3} \ln(\Lambda)\,G^2\,n_{\ast}(m_{\rm bh}+m_{\ast})^2
\label{eq.Gammarlx}
\end{equation}

\begin{figure}
\resizebox{\hsize}{!}
          {\includegraphics[scale=1,clip]{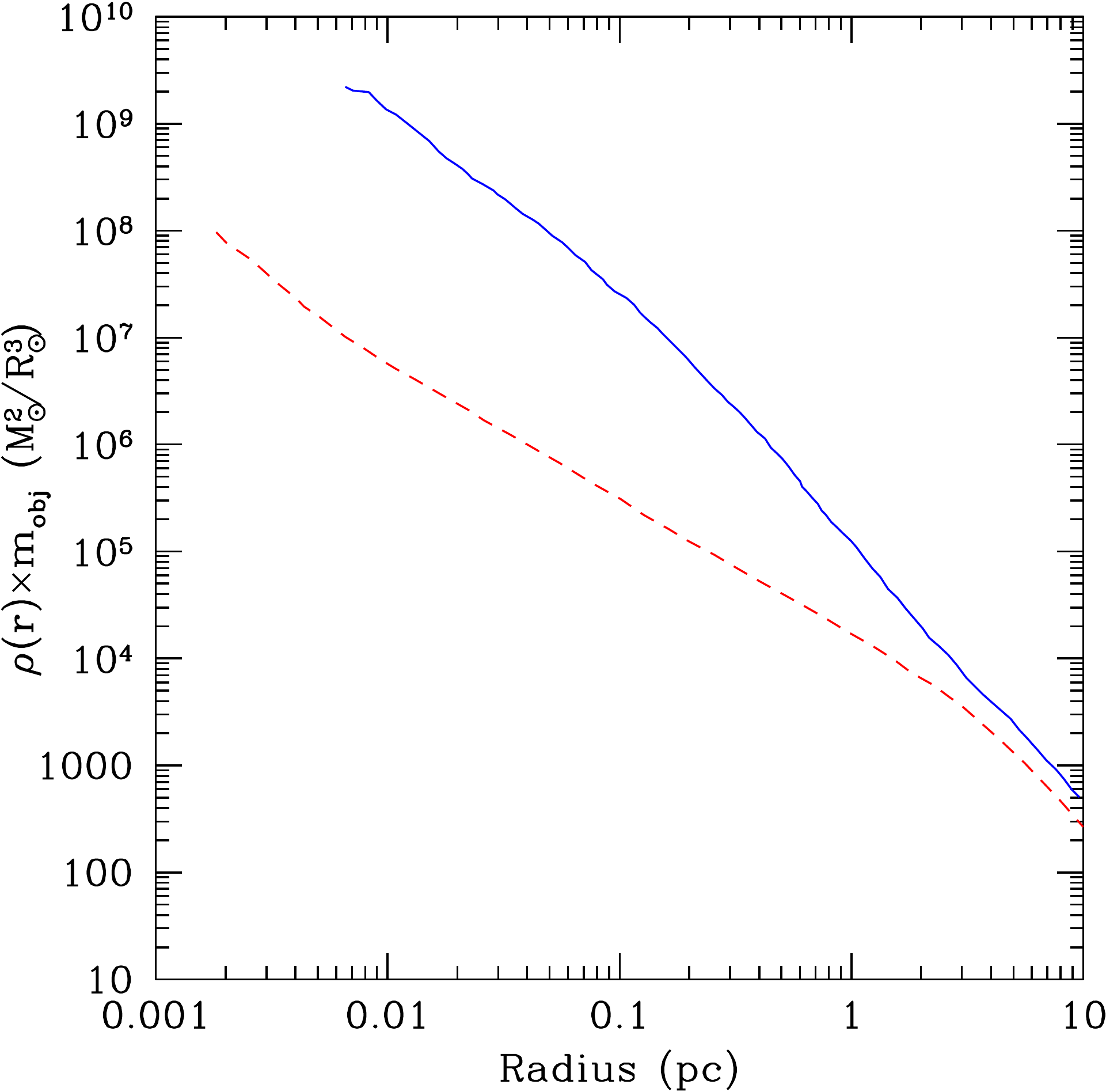}}
\caption
   {
Relaxation rate for the Milky-Way model {\tt GN25} of \cite{FAK06a} in $M_{\odot}^2$
per volume as a function of the radius from SgrA*. The red, dashed
line corresponds to MS stars and the blue,
solid line to stellar-mass black holes.
   }
\label{fig.InverseTrlx}
\end{figure}

This last equation expresses the ``encounter relaxation time'', which depends
on the characteristics of a peculiar class of encounter (see
\cite{Amaro-SeoaneLRR2012}).  I.e., for our purposes, a stellar-mass black hole
of mass $m_{\rm bh}$ with a field MS star of mass $m_{\ast}$ with a local
density $n_{\ast}$ and a relative velocity $v_{\rm rel}$.  We can depict
Eq.~(\ref{eq.Gammarlx}) thanks to the Monte-Carlo simulations of the Galactic
Centre of \cite{FAK06a}. In their Fig.~10, right panel, they give the evolution
of the density $\rho(r)$ profile for a standard Milky-Way nucleus after
$1.05\times 10^{10}\,{\rm yrs}^{-1}$. From Eq.(\ref{eq.Gammarlx}), $\Gamma_{\rm
rlx} \propto n(r)\times m_{\rm obj}^2$, i.e. $\Gamma_{\rm rlx} \propto \rho(r)
\times m_{\rm obj}$, with $m_{\rm obj}$ the mass of the object taken into
consideration, a BD or a stellar-mass black hole. In
Fig.~(\ref{fig.InverseTrlx}) we show this quantity from the data of
\cite{FAK06a}. We can see that from a distance of about $10\,{\rm pc}$, the
relaxation rate is dominated by stellar-mass black holes. We note that the fact
that within $\lessapprox 0.1~{\rm pc}$ the stellar-mass black holes have a
tendency to follow a distribution which looks shallower than the one from the
MS stars is due to a problem related to the resolution of the Monte Carlo
simulations (M. Freitag, personal communication). On the other hand, we note
that by assuming a pure power-law, as we are doing in our analytical approach,
we are artificially increasing $T_{\rm rlx}$, since we are populating with more
stellar-mass black holes the innermost radii. Therefore, the results that we
will derive for Eq.~(\ref{eq.EventRate}) are to be regarded as a lower limit.

Secondly (ii), it must be noted that by assuming that relaxation is dominated
by stellar-mass black holes, we are implying that relaxation can be added up
individually from two mass groups, BDs and stellar-mass black holes, and that
the contribution from BDs is negligible. In star cluster evolution, close to
the central regions, energy equipartition is found only among the largest
masses, and it progressively moves towards velocity equipartition at low masses
(see e.g.  \cite{BianchiniEtAl2016}). Hence, if the distribution function of
mass and velocity is $f(m,\,v)$ with $v$ the velocity and $m$ the mass, and a
moment of the change of velocities is of the form

\begin{equation}
<dv^2> = \int dv^2 f(m,\,v)\, dm\, dv,\nonumber
\end{equation}

\noindent
since energy equipartition among the low-mass object can be neglected, this
last equation can be expressed as

\begin{equation}
<dv^2> = \sum_m n(m) \left(\int dv^2 f(v) dv\right), \nonumber
\end{equation}

\noindent
with $n(m)$ the density of stars of mass $m$.  This has two important
implications. First, we expect BDs to actually be close to the centre and,
secondly, since the mass of the BD population is only a small contribution to
the relaxation produced by stellar-mass black holes, we ignore their
contribution in the calculations related to relaxational processes.

\subsection{The critical and minimum radii}

We now need to calcuate the only remaining quantities, (4) the critical and the
minimum semi-major axis. The critical radius, the upper limit of integral
Eq.~(\ref{eq.EventRate}), can be derived by taking into account its definition.
This is the semi-major axis at which the threshold
curve which separates the dynamics- and GW-regime merges with the LSO curve. First, we equate the relaxation time
to the inspiral timescale for a binary made from the MBH and a BD,

\begin{equation}
T_{\rm rlx,\,peri } = C\,T_{\rm GW}(a,\,e)
\label{eq.TrlxTGW}
\end{equation}

\noindent
In this equation, $T_{\rm rlx,\,peri}$ is the relaxation time at pericentre,
i.e., $T_{\rm rlx,\,peri }:=T_{\rm rlx}(a)\times (1-e)$ \citep[][]{Amaro-SeoaneLRR}, and $C \sim 1$.
Since at the radii of interest the driving species in relaxation is that of stellar-mass
black holes, the mass that matters in the expression of $T_{\rm rlx,\,peri}$ is
$m_{\rm bh}$. However, we are interested in the inspiral of a BD into the MBH,
and hence the mass which is relevant for the right-hand side is $m_{\rm BD}$.
We assume here that $e \sim 1$, which is the characteristic eccentricity of
EMRIs when they form \citep[][]{Amaro-SeoaneLRR}, so that the function
$f(e)$ which appears in $T_{\rm GW}(a,\,e)$, \cite{Peters64}-, can be
approximated as $f(e) = 425/(768\,\sqrt{2})$.

{We assume Newtonian parabolic orbits because BDs will have semi-major
axes much larger than their pericentre distance. We hence equate the Newtonian
value of the pericentre distance of the last stable parabolic orbit around the
massive black hole \citep{ST83} to the pericentre distance,}

\begin{equation}
\frac{8\,G{M}_{\rm BH}}{c^2} = a\,(1-e) \cal{W}(\iota,\,{\rm s}).
\label{eq.PeriLSO}
\end{equation}

\noindent
In this equation, we have multiplied the right-hand side by the function
$\cal{W}(\iota,\,{\rm s})$, which takes into account the impact of the
asymmetry between prograde and retrograde orbits on the location of the LSO for
a Kerr MBH with respect to the Schwarzschild case (at a distance $4\,R_{\rm S}$)
\cite{Amaro-SeoaneSopuertaFreitag2013}.  This function depends on the
inclination of the orbit, $\iota$ and the magnitude of the spin of the MBH,
$\rm{s}$.

Therefore, we obtain that

\begin{equation}
T^{}_{\rm GW}(a,\,e) \sim \sqrt{2}\,\frac{24}{85}\frac{c^5}{G^3}
                        \frac{a^4\,\left(1-e\right)^{7/2}}{m^{}_{\rm BD}\,M_{\rm BH}^2}.
\label{eq.Tgwae}
\end{equation}

\begin{figure}
\resizebox{\hsize}{!}
          {\includegraphics[scale=1,clip]{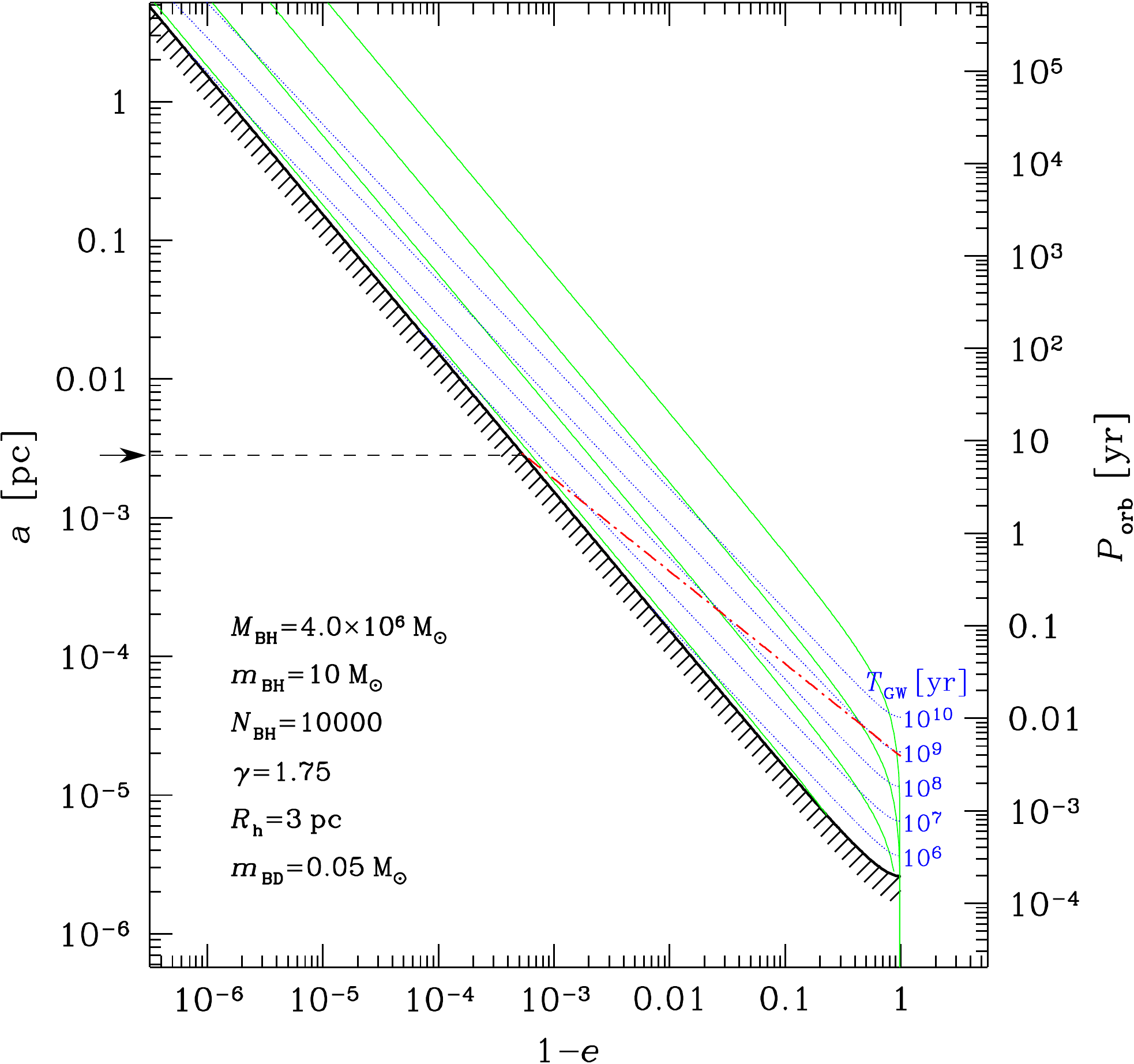}}
\caption
   {
Definition of the critical radius $a_{\rm crit}$ in the
phase space in the semi-major axis (in parsecs) and eccentricity plane,
$a$\,--\,$(1-e)$, for the inspiral of a BD of mass $m_{\rm
BD}=0.05\,M_{\odot}$, into a MBH of mass $4\times 10^6\,M_{\odot}$. The dashed,
blue lines are isochrones depicting the inspiral time $T_{\rm GW}$ in years of
the binary, were this to evolve only due to the emission of gravitational
radiation, as estimated in the approximation of \cite{PM63}. When a
system crosses one of the isochrones, it will inspiral in a time as
shown by the corresponding curves. The green curves show the relation between
$a$ and $e$ as estimated in the work of \cite{Peters64}, again under the same
assumption. The red, dashed line gives the threshold for two different regimes
in the evolution of the binary, as derived in Eq.~(\ref{eq.1_e_a}).
Above this line the binary evolves due to
two-body relaxation, while below the curve the driving mechanism is the
emission of GWs. The solid, black line crossing the figure from the
top left to the bottom right is the last stable orbit (LSO) for a Schwarzschild
MBH of that mass. The conjunction of the red line with the LSO defines
the critical semi-major axis $a_{\rm crit}$, which is shown with an arrow
on the y-axis.
   }
\label{fig.a_crit}
\end{figure}

From Eqs.~(\ref{eq.TrlxTGW}), (\ref{eq.Tr}) and~(\ref{eq.t0full}), we can obtain the
relation between $a$ and $e$, {the threshold curve between the dynamics-dominated regime
and the gravitational-wave one, which we use for the dashed, red curve of Fig.(\ref{fig.a_crit}),}

\begin{align}
\left( 1-e \right)^{5/2} & = \frac{4.26}{(3-\gamma)(1+\gamma)^{3/2}}\frac{85}{24}
                           \frac{1}{\sqrt{2}\,c^5}\frac{G^{\,5/2}}{\ln(\Lambda)}\nonumber\\
                         & \frac{M_{\rm BH}^{7/2}}{N_0}
                           \frac{m_{\rm BD}}{m_{\rm bh}}
                           R_0^{3-\gamma}a^{\gamma-11/2}
\label{eq.1_e_a}
\end{align}

\noindent
As before, $R_0$ is a characteristic radius, within which relaxation is
dominated by stellar-mass black holes, and $N_0$ is the number of them
contained in that radius. Because of the explanation we gave before about 
stellar-mass black holes dominating relaxation, we choose now to set this
radius also to the influence radius, $R_{\rm h}$, as we did with the normalisation
of the BDs and MS stars.

We can see that the threshold for
these binaries to decouple from the stellar system is around a distance of
$2.5\times 10^{-3}\,{\rm pc}$.  Indeed, solving the same equations for
$a^{}_{\rm crit}$, we have that

\begin{align}
a_{\rm crit} &=  \epsilon \, R_{0} \times \nonumber \\
               &                   \times  \left[
                                           {\cal W}(\iota,\,{\rm s})^{5/2}\,N_{0} \ln(\Lambda)
                                           \left(\frac{M_{\rm BH}}{m_{\rm BD}}\right)
                                           \left(\frac{M_{\rm BH}}{m_{\rm bh}}\right)^{-2}
                                            \right]^{\frac{1}{\gamma-3}},
\label{eq.acrit}
\end{align}

\noindent
where we have defined

\begin{equation}
 \epsilon := \left[\frac{C}{4.26}\frac{6144}{85}(3-\gamma)(1+\gamma)^{3/2} \right]^{\frac{1}{\gamma-3}}.
\label{eq.epsilon}
\end{equation}

Adopting $\gamma=1.75$, $R_0=R_{\rm h}=3\,{\rm pc}$, we have that BDs have the following $a_{\rm crit}$ at the GC

\begin{align}
a_{\rm crit}  & \sim 2.89\times10^{-3}\,{\rm pc}\,{\cal W}(\iota,\,{\rm s})^{-2}\times \nonumber \\
              &\left(\frac{M_{\rm BH}}{4\cdot 10^6 M_{\odot}}\right)^{4/5}
               \left(\frac{m_{\rm BD}}{0.05 M_{\odot}}\right)^{4/5} \times \nonumber \\
              & \left(\frac{m_{\rm bh}}{10 M_{\odot}}\right)^{-8/5}
                    \left(\frac{N_{0}}{10^4}\right)^{-4/5}\left(\frac{ln(\Lambda)}{12}\right)^{-4/5}.
\label{eq.acritBD}
\end{align}

\noindent
In Fig.~(\ref{fig.a_crit}) we depict this threshold for the same values in
phase-space for the values of $\gamma$ and $\beta$ of \citep{BW77}.  We note that our analytical model does not take into account the
fact that dynamical friction will bring in more stellar-mass black holes within
the influence radius. For instance, the Milky-Way models of \cite{FAK06a} find
$\sim 2\times10^4$ stellar-mass black holes at a distance of
$R=1\,\textrm{pc}$. However, since the dependency of $a_{\rm crit}$ on $N_0$
has an exponent of $4/5$, the difference is small. This is also true for the dependency on $m_{\rm
BD}$, with e.g. $a_{\rm crit} \sim 5\times10^{-3}\,{\rm pc}\,{\cal W}(\iota,\,{\rm s})^{-2}$
for $m_{\rm BD}=0.1\,M_{\odot}$. 

The minimum radius is the distance within which we expect to have at least one object to start the
integration. This can be derived from Eq.~(\ref{eq.Nsubpop}) taking into account that 
$N^{\rm BD}_{0\,\textrm{MS}}={M_{\rm BH}}/{\bar{m}_{\ast}}$. Hence,

\begin{equation}
a_{\rm min} \simeq 1.65\times 10^{-5}\,\textrm{pc}\,f_{\ast,\,\textrm{sub}}^{-2/3}\left(\frac{R_{\rm h}}{1\,\textrm{pc}}\right),
\end{equation}

\noindent
with $f_{\ast,\,\textrm{sub}}$ the fraction of substellar objects or stars taken into consideration, and adopting
$\beta=3/2$.

With all of the quantities defined, we substitute them into Eq.~(\ref{eq.EventRate}) and find that the integral
has as solution

\begin{align}
&\dot{\Gamma}_{\rm X-MRI} =  \frac{3-\beta}{2\,\lambda}\frac{N^{\rm BD}_{0\,\textrm{MS}}}
                                                {T_0\,R_{\rm h}^{\lambda}} f_{\rm sub}^{\rm BD} \times \nonumber \\
                                  &
                                  \left\{
                                       a_{\rm crit}^{\lambda}\left[\ln(\Lambda_{\rm crit}) -\frac{1}{\lambda}\right] -
                                       a_{\rm min}^{\lambda}\left[\ln(\Lambda_{\rm min}) -\frac{1}{\lambda}\right]
                                  \right\},
\label{eq.FinalGammaBD}
\end{align}

\noindent
where we have introduced

\begin{align}
\lambda            & := 9/2-\beta-\gamma \nonumber \\
\Lambda_{\rm crit} & := \left( \frac{a_{\rm crit}}{8\,R_{\rm S}} \right) \nonumber \\
\Lambda_{\rm min} & := \left( \frac{a_{\rm min}}{8\,R_{\rm S}} \right).
\end{align}

\subsection{Results for two classical examples}

We now give two examples for the values of the critical parameters for two
different solutions of the power-laws.  For historical reasons, we use the
values derived by \cite{BW77}, but note that the values found by the authors,
$\gamma=7/4$ and $\beta \rightarrow 3/2$ are a (heuristically) generalised
solution of their earlier work \cite{BW76} that only depends on the mass ratio
of the two different populations. While this solution is mathematically
correct, it assumes a stellar population in which 50\% of all stars are
stellar-mass black holes. We refer to this model using the sub- and superscript
``BW''.

We then give a more accurate value, which corresponds to more appropriate
number fractions. This translates into a more efficient diffusion, as noted by
\cite{AlexanderHopman09} and \cite{PretoAmaroSeoane10}.  In particular, $\gamma
= 2 $ and the population of stars with lighter masses $\beta=3/4$, as found
with the direct-summation $N-$body simulations of \cite{PretoAmaroSeoane10}
(see also \cite{Amaro-SeoanePreto11}). The notation in this case is ``SMS''
(i.e. strong-mass segregation, a term coined by Tal Alexander).

For legibility, we now introduce the following notation for standard values of normalisation,

\begin{align}
\tilde{\Lambda}& :=\left(\frac{\ln(\Lambda)}{13}\right),~ \tilde{N}_{0}:=\left(\frac{N_0}{12000}\right)\nonumber \\
\tilde{R}_{0}& :=\left(\frac{R_{\rm h}}{1\textrm{pc}}\right),~ \tilde{m}_{\rm BD}:=\left(\frac{m_{\rm BD}}{0.05\,M_{\odot}}\right).
\end{align}

We have chosen the value of $\tilde{\Lambda}$ based on the fact that the galactic nucleus is a non self-gravitating
system, so that $\ln(\Lambda) \simeq \ln(M_{\rm BH}/m_{\rm bh}) \sim 12.9$ (see e.g. \cite{Amaro-SeoaneLRR2012}).

The BW case leads to the following results for the critical radius and the normalization timescale $T_0$,

\begin{align}
a^{\rm BW}_{\rm crit} & \sim 8\times 10^{-4}\,\textrm{pc}\,{\cal W}(\iota,\,{\rm s})\tilde{R}_{0} \tilde{N}_{0}^{-4/5}\tilde{\Lambda}^{-4/5}\tilde{m}_{\rm BD}^{4/5}\nonumber\\
T^{\rm BW}_0 & \sim 10.34 \times 10^9\,\textrm{yrs} \, \tilde{R}_{0}^{3/2}\tilde{N}_{0}^{-1}\tilde{\Lambda}^{-1}.
\end{align}

Hence, the BW event rate is

\begin{align}
\dot{\Gamma}_{\rm BW} & \sim 1.8\times 10^{-4}\,\textrm{yrs}^{-1}\tilde{N}_{0}\,\tilde{\Lambda}\,\tilde{R}_{0}^{-11/4} \times \nonumber\\
                           & \Bigg\{
                                    1.34\times 10^{-4} \tilde{R}_{0}^{5/4}\tilde{N}_{0}^{-1}\tilde{\Lambda}^{-1}\tilde{m}_{\rm BD}\,{\cal W}(\iota,\,{\rm s})\times \nonumber\\
                           &         \left[\ln\left(262\, \tilde{R}_{0}\, \tilde{N}_{0}^{-4/5} \tilde{\Lambda}^{-4/5}\tilde{m}_{\rm BD}^{4/5}\,{\cal W}(\iota,\,{\rm s})^{-2}\right) - \frac{4}{5} \right] - \nonumber \\
                           &         6.86\times 10^{-25/4}\tilde{R}_{0}^{5/4}\times \left[\ln\left(15.22\, \tilde{R}_{0}\right) - \frac{4}{5} \right] 
                             \Bigg\}.
\end{align}

The same quantities calculated for the SMS case are,

\begin{align}
a^{\rm SMS}_{\rm crit} \sim & 1.4\times 10^{-4}\,\textrm{pc}\,{\cal W}(\iota,\,{\rm s})^{-5/4} \tilde{R}_{0} \tilde{N}_{0}^{-1} \tilde{\Lambda}^{-1}\tilde{m}_{\rm BD}\nonumber\\
T^{\rm SMS}_0 \sim & 1.13 \times 10^9\,\textrm{yrs} \, \tilde{R}_{0}^{3/2}\tilde{N}_{0}^{-1}\tilde{\Lambda}^{-1}.
\end{align}

With these, the event SMS rate can be derived to be

\begin{align}
\dot{\Gamma}_{\rm SMS} & \sim 2.3\times 10^{-3}\,\textrm{yrs}^{-1} \tilde{R}_{0}^{-5/2}\tilde{N}_{0}\tilde{\Lambda} \times \nonumber\\
                           & \Bigg\{
                                    1.4\times 10^{-4} \tilde{R}_{0}\, \tilde{N}_{0}^{-1} \tilde{\Lambda}^{-1} \tilde{m}_{\rm BD}\,{\cal W}(\iota,\,{\rm s})^{-5/2}\times \nonumber\\
                           &         \left[\ln\left(46\,\tilde{R}_{0}\, \tilde{N}_{0}^{-1}\tilde{\Lambda}^{-1}\tilde{m}_{\rm BD}\,{\cal W}(\iota,\,{\rm s})^{-5/2}\right) - 1  \right] - \nonumber \\
                           &         4.67\times 10^{-7} \tilde{R}_{0} \times \left[\ln\left(15.24\, \tilde{R}_{0}\right) - 1\right]
                             \Bigg\}.
\end{align}

\begin{figure*}
\resizebox{\hsize}{!}
          {\includegraphics[scale=1,clip]{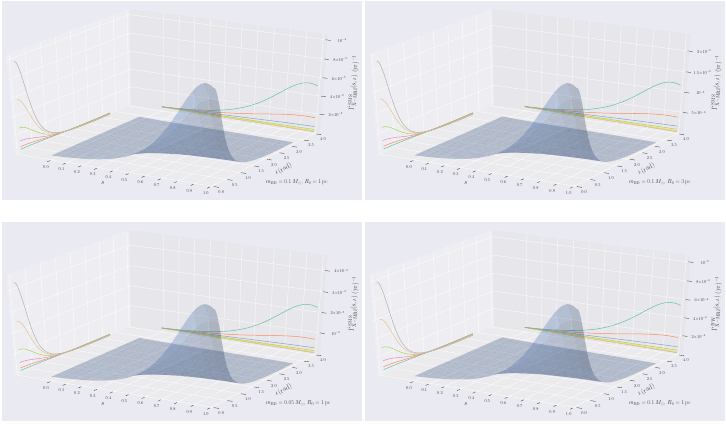}}
\caption
   {
Number of sources per year at the GC that successfully inspiral towards SgrA*
as a function of the spin of the MBH and the inclination of the orbit $\iota$
in rad, for typical values of the mass of the BD. We include as well two
different values of the influence radius, which is the value we have chosen for
the normalization radius $R_0$, of $1\,\textrm{pc}$ (for being a traditional
value in the related literature) and of $3\,\textrm{pc}$ (according to more
recent observations, see \citep{SchoedelEtAl2014,SchoedelEtAl2018}).  We show
three characteristic combinations for the SMS scenario and one for the less
realistic BW case. 
   }
\label{fig.Gamma_BD}
\end{figure*}

In Fig.~(\ref{fig.Gamma_BD}) we show a few examples for the SMS cases,
including one for the BW.  The rates typically are about
$10^{-5}\,\textrm{yr}^{-1}$ for the SMS case and one order of magnitude less
for the less realistic BW scenario.

\subsection{A check of our model}

Thanks to Eqs.~(\ref{eq.FinalGammaBD}, \ref{eq.acrit}, \ref{eq.t0full}), we can
evaluate our results by simplyfing our analysis to the specific case of
stellar-mass black holes. We can easily re-calculate the previous equations for
this kind of stellar objects and define the quantity $\tilde{m}_{\rm bh} :=
m_{\rm bh}/(10\,M_{\odot})$ by analogy with the previous section. We find that
the event rate for the BW case is

\begin{align}
\dot{\Gamma}_{\rm BW,\,bh} & \sim 2.63\times 10^{-6}\,\textrm{yrs}^{-1}\tilde{N}_{0}\,\tilde{\Lambda}\,\tilde{R}_{0}^{-5/2}\,\tilde{m}_{\rm bh}^2 \times \nonumber\\
                           & \Bigg\{
                                    5\times 10^{-2} \tilde{R}_{0}\tilde{N}_{0}^{-4/5}\tilde{\Lambda}^{-4/5}\tilde{m}_{\rm bh}^{4/5}\,{\cal W}(\iota,\,{\rm s})^{-2}\times \nonumber\\
                           &         \left[\ln\left(16318\, \tilde{R}_{0}\, \tilde{N}_{0}^{-4/5} \tilde{\Lambda}^{-4/5}\tilde{m}_{\rm bh}^{4/5}\,{\cal W}(\iota,\,{\rm s})^{-2}\right) - 1 \right] \times \nonumber \\ 
                           &  2\times 10^{-3}\tilde{R}_{0}\times \left[\ln\left(618\, \tilde{R}_{0}\right) - 1 \right]
                             \Bigg\}.
\end{align}

For the standard value of $\tilde{m}_{\rm bh}=1$ (and the rest of parameters
set to unity), we recover the usual rate of $\dot{\Gamma}_{\rm BW,\,bh}\sim
10^{-6}\,\textrm{yr}^{-1}$ \citep[see e.g.][and references therein]{Amaro-SeoaneLRR2012}.  
For more massive stellar-mass black holes,
in particular for $\tilde{m}_{\rm bh}=4$ (i.e. $m_{\rm bh}=40\,M_{\odot}$),
which is the scenario recently proposed by \citep{EmamiLoeb2019} for EMRIs at
the GC, we find a slightly enhanced rate, but still negligible,
$\dot{\Gamma}_{\rm BW,\,bh}\sim 6.2\times10^{-5}\,\textrm{yr}^{-1}$.

For completness, we give now the case corresponding to SMS,

\begin{align}
\dot{\Gamma}_{\rm SMS,\,bh} & \sim 1.92\times 10^{-6}\,\textrm{yrs}^{-1}\tilde{N}_{0}\,\tilde{\Lambda}\,\tilde{R}_{0}^{-2}\,\tilde{m}_{\rm bh}^2 \times \nonumber\\
                           & \Bigg\{
                                    1.6\times 10^{-1} \tilde{R}_{0}^{1/2}\tilde{N}_{0}^{-1/2}\tilde{\Lambda}^{-1/2}\tilde{m}_{\rm bh}^{1/2}\,{\cal W}(\iota,\,{\rm s})^{-5/4}\times \nonumber\\
                           &         \left[\ln\left(9138\, \tilde{R}_{0}\, \tilde{N}_{0}^{-1} \tilde{\Lambda}^{-1}\tilde{m}_{\rm bh}\,{\cal W}(\iota,\,{\rm s})^{-5/2}\right) - 2 \right] - \nonumber \\
                           &         4\times 10^{-2}\tilde{R}_{0}^{1/2}\times \left[\ln\left(618\, \tilde{R}_{0}\right) - 2 \right]
                             \Bigg\}.
\end{align}

\section{Number of sources in band}

Contrary to EMRIs, which are more massive, X-MRIs spend a long time in band,
because they undergo $\sim\,10^8$ cycles before they cross the event horizon of
the MBH.  In Fig.~(\ref{fig.BD_SNR}) we see that they can spend as much as
$\sim 10^6\,{\rm yr^{-1}}$ with a SNR $>10$.  Since the event rate at the GC is
of about $\dot{\Gamma}_{\rm X-MRI}\sim 10^{-5}\,{\rm yr}^{-1}$, we could
naively argue that at any given time there should be $\sim 10 \times
\dot{\Gamma}_{\rm X-MRI}$ sources at different frequencies. These would have
different SNRs, from tens to a few $10^3$ or even $\gtrsim 10^4$, depending on
their mass.

\begin{figure}
\resizebox{\hsize}{!}
          {\includegraphics[scale=1,clip]{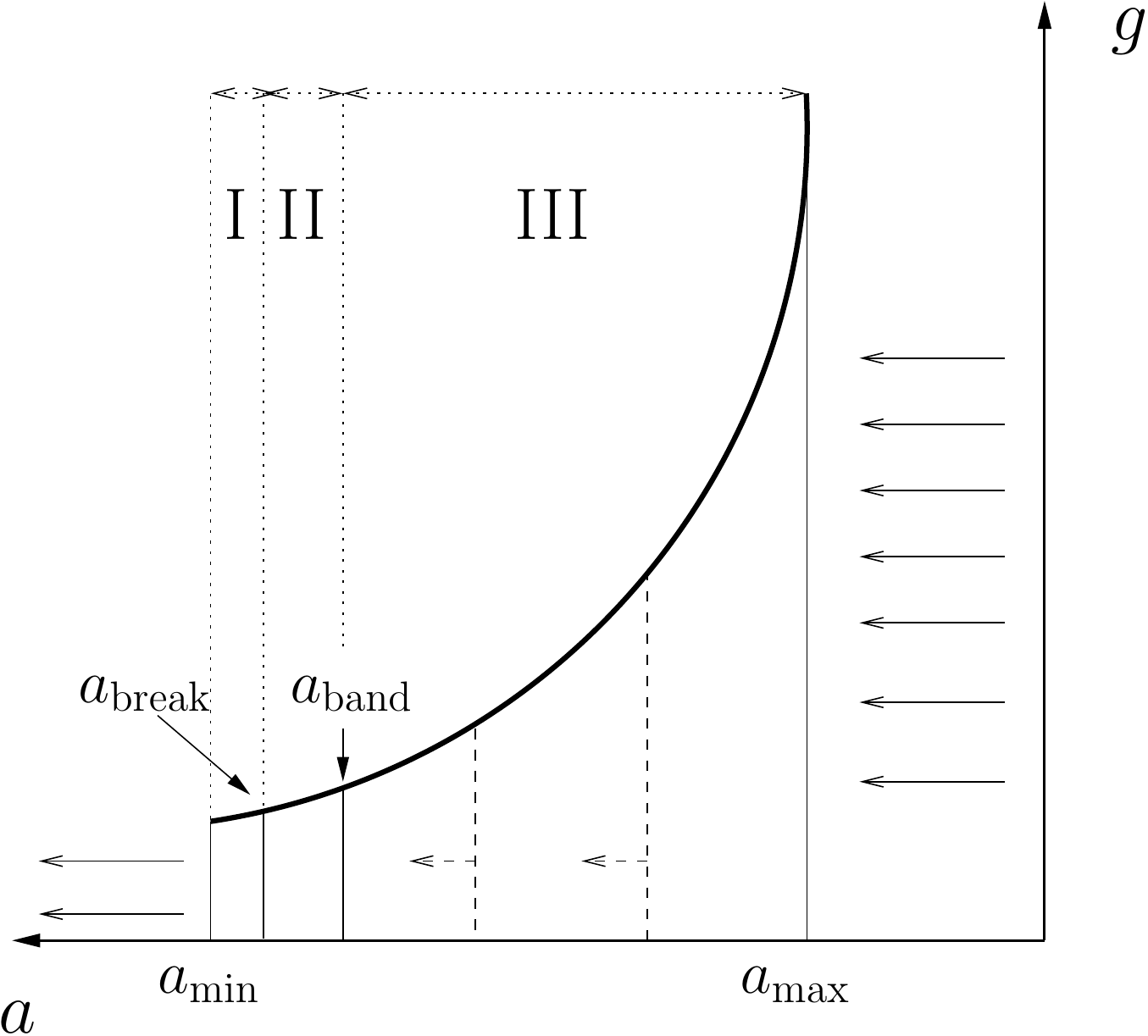}}
\caption
   {
Density of sources $g=dN/da$ as a function of $a$.  Sources come from larger
values of $a$, up to a maximum value at $a_{\rm max}$ ---which is equivalent to
$a_{\rm crit}$--- and diffuse towards lower values-, down to a minimum $a_{\rm min}$.  Sources with
semi-major axis values above $a_{\rm band}$ emit GWs but at too low frequencies
for detection. Those that cross the threshold value are in
band. Sources above $a_{\rm break}$ have large
eccentricities, while sources below it can be regarded as circular. In the
text we derive the amount of sources in the regimes marked as ${\rm I}$ (with
semi-major axis values between $a_{\rm min}$ and $a_{\rm break}$), ${\rm II}$
(between $a_{\rm break}$ and $a_{\rm band}$) and ${\rm III}$ (between $a_{\rm band}$
and $a_{\rm max}$).
   }
\label{fig.g_a_xfig}
\end{figure}

However, not all of these objects will be successful X-MRIs, because
their semi-major axis must be below a threshold value, which we henceforth call
$a_{\rm band}$.  To derive this value and, therefore, the number of sources at
any given time, we need to evaluate the number of sources $N$ at any given
semi-major axis $a$; i.e. we need to assess the (line) density function

\begin{equation}
g = \frac{dN}{da}
\label{eq.g}
\end{equation}

\noindent
as a function of $a$. In Fig.~(\ref{fig.g_a_xfig}) we depict an illustrative shape of $g$.
Because of the slow diffusion, $g$ must have larger values at larger values of $a$. However,
the number of sources sweeping the different range of values for $a$ is constant, so that
$\Delta N(t=\Delta t)=\Delta N(t=0)$. This is equivalent to

\begin{equation}
\Delta t \left(v_rg_r - v_lg_l \right) = - \left( g(\Delta t)\Delta a - g(0) \Delta a  \right),
\end{equation}

\noindent
where $v_l$, $g_l$ (and $v_r$, $g_r$) are respectively the values of the
velocity and density function of the left (right)-, dashed line of
Fig.~(\ref{fig.g_a_xfig}). $g(0)$ is the density function at time $t=0$- and
$g(\Delta t)$ after a time $\Delta t$, where the negative sign accounts for the fact
that we are losing sources when crossing $a_{\rm min}$. Therefore,

\begin{equation}
\frac{v_rg_r - v_lg_l}{\Delta a} = - \frac{g(\Delta t) - g(0)}{\Delta t}.
\end{equation}

In the limit $\Delta t \to 0$, this can be written
as the continuity equation

\begin{equation}
\frac{\partial}{\partial a} \left( \dot{a}(a,\,e) g \right) + \frac{\partial g}{\partial t} =0,
\label{eq.cont}
\end{equation}

\noindent
where $\dot{a}(a,\,e)$ is the velocity, and it is a function of the
semi-major axis $a$ and the eccentricity $e$. Its explicit form can be found in
\cite{Peters64} in the approximation of Keplerian ellipses,

\begin{align}
\dot{a}(a,\,e) & = - \frac{64}{5} \frac
                                {G^3\,M_{\rm BH}m_{\rm BD}(M_{\rm BH}+m_{\rm BD})}
                                {c^5a^3(1-e^2)^{7/2}} \nonumber \\
               &                \left(
                                1 + \frac{73}{24}e^2 + \frac{37}{96}e^4
                                \right),
\label{eq.adot}
\end{align}

\noindent
Eq.~(\ref{eq.cont}) states that the rate at which sources enter the range of
semi-major axis values is equal to the rate of sources leaving the system plus
their accumulation.  Since the density function does not vary in time,
the right term vanishes. After integrating we obtain

\begin{equation}
\dot{a}(a,\,e)\,g = K,
\label{eq.DgK}
\end{equation}

\noindent
with K a constant. In Fig.~(\ref{fig.g_a_xfig}) we display a representative
illustration of $a_{\rm band}$.  The total number of sources is the amount of
inspiraling BDs we derived in Sec.~(\ref{sec.Crossing}), and are comprised
between the values $a_{\rm max}$ and $a_{\rm min}$, so that all we need to do
is to obtain the relative occupation fractions in the two areas to derive the
number of sources detectable by LISA, i.e., those with semi-major axis between
$a_{\rm band}$ and $a_{\rm min}$.

From Eqs.~(\ref{eq.g},\,\ref{eq.DgK}), we have that

\begin{equation}
\frac{dN}{da}=\frac{K}{\dot{a}(a,\,e)}.
\end{equation}

\noindent
In order to integrate this function, we must distinguish two different regimes.
Those X-MRIs with semi-major axis values above a threshold $a_{\rm break}$
still have a significant amount of eccentricity, while for lower values of the
semi-major axis, the systems will be circular, or close to circular.  We first
address the eccentric regime. In Eq.~(\ref{eq.adot}) we now
use the fact that the pericentre distance is $r_p=a(1-e)$. Therefore,

\begin{equation}
dN = Z(e) r_p^{7/2}a^{-1/2}da,
\end{equation}

\noindent
where we have introduced

\begin{equation}
Z(e) \equiv -\frac{5K}{64}\frac{c^5}{G^3}\frac{L(e)}{M_{\rm BH}m_{\rm BD}(M_{\rm BH}+m_{\rm BD})},
\end{equation}

\noindent
and

\begin{equation}
L(e):= \frac{\left(1-e\right)^{7/2}}{\left( 1 + \frac{73}{24}e^2 + \frac{37}{96}e^4 \right)}.
\end{equation}

Since we are considering high values of eccentricity, $r_p$ is constant to first order (as can be seen in Fig.~(\ref{fig.isochr})) and can be taken out of the integral. In addition, $e \sim 1$ so that L(e) is constant and can also be taken out of the integral. Therefore, the number of sources between $a_{\rm max}$ and $a_{\rm break}$ is

\begin{equation}
\frac{dN}{da}\Big|_{a>a_{\rm break}} = \frac{dN}{da}\Big|_{a_{\rm break}} \left(\frac{a}{a_{\rm break}}\right)^{-1/2},
\label{eq.dNdaBigecc}
\end{equation}

\noindent
where we have normalized the distribution to $a_{\rm break}$. Integrating,

\begin{align}
N(a>a_{\rm break}) & = 2 \times a_{\rm break} \frac{dN}{da}\Big|_{a_{\rm break}} \nonumber \\
                   & \left[ \left( \frac{a}{a_{\rm break}} \right)^{1/2} \right]_{a_{\rm break}}^{a_{\rm max}}.
\end{align}

For $a<a_{\rm break}$, we need to take into account that $e \sim 0$ in Eq.(\ref{eq.adot}), so that

\begin{equation}
\frac{dN}{da}\Big|_{a<a_{\rm break}} = \frac{dN}{da}\Big|_{a_{\rm break}} \left(\frac{a}{a_{\rm break}}\right)^{3},
\end{equation}

\noindent
and, as in Eq.~(\ref{eq.dNdaBigecc}) we have normalized to the conjunction, $a_{\rm break}$. Therefore,

\begin{align}
N(a<a_{\rm break}) & = \frac{1}{4} \times a_{\rm break} \frac{dN}{da}\Big|_{a_{\rm break}} \nonumber \\
                   & \left[ \left( \frac{a}{a_{\rm break}} \right)^{4} \right]_{a_{\rm break}}^{a_{\rm min}}.
\label{eq.dNdaLowecc}
\end{align}

By using Eq.~(\ref{eq.dNdaBigecc}), we obtain the ratios for the number of
sources in ${\rm II}$ and ${\rm III}$ of Fig.~(\ref{fig.g_a_xfig}),

\begin{equation}
\frac{N_{\rm II}}{N_{\rm III}}  = \frac{a_{\rm band}^{1/2}-a_{\rm break}^{1/2}}{a_{\rm max}^{1/2}-a_{\rm band}^{1/2}}.
\end{equation}

Likewise, by using Eqs.~(\ref{eq.dNdaBigecc}) and (\ref{eq.dNdaLowecc}), we
can obtain the ratio of sources between the region ${\rm I}$ (i.e., $N(a<a_{\rm
break})$) and ${\rm II}+{\rm III}$ (i.e., $N(a>a_{\rm break})$),

\begin{equation}
\frac{N_{\rm I}}{N_{\rm II}+N_{\rm III}} = \frac{1}{8} \times \frac{1-\left(a_{\rm min}/a_{\rm break}\right)^4}{\left(a_{\rm max}/a_{\rm break} \right)^{1/2}-1}.
\end{equation}

Finally, as we have discussed at the beginning of this section, we know that
the total amount of sources in ${\rm I}$, ${\rm II}$ is
Eq.~(\ref{eq.FinalGammaBD}) multiplied by the typical lifetime of these sources
at a given eccentricity with a minimum SNR of 10, $T\left(a_{\rm max},e\right)
\sim 2\times 10^6~{\rm yr}^{-1}$,

\begin{equation}
N_{\rm I}+N_{\rm II} = \dot{\Gamma}_{\rm X-MRI} \times T\left(a_{\rm max},e\right).
\end{equation}

\noindent
The plunge radius is $a_{\rm min} = 2 \times R_{\rm S} \sim 7.67\times
10^{-7}~{\rm pc}$.  From Fig.~(\ref{fig.isochr}), we see that a representative value of $a_{\rm break}
\sim 4 R_{\rm S} \sim 1.53\times 10^{-6}~{\rm pc}$.  The value of $a_{\rm
max}$ is given by Eq.~(\ref{eq.acritBD}), and a representative value of $a_{\rm band}$ with SNR=$10$ can
be directly read from Fig.~(\ref{fig.isochr}).
Therefore, we obtain that

\begin{align}
N_{\rm I}   & \sim 5   \nonumber \\
N_{\rm II}  & \sim 15.
\end{align}

We note that these values fluctuate by a multiplying factor of a few depending
on the location of $a_{\rm band}$, which depends on the initial eccentricity of
the source.

\section{Conclusions}

Brown dwarfs can inspiral in our Galactic Centre on to SgrA* via the emission
of gravitational waves without suffering significant tidal stresses.

Because they have a mass ratio of $q\sim 10^8$, the time these systems spend in
the LISA band is very long, since the number of times they revolve around the
MBH is proportional to $q$. These systems, which we call X-MRIs, start to
accumulate a signal-to-noise ratio of SNR=$10$ some $\sim 2\times 10^6~{\rm
yrs}$ before the final plunge through the event horizon. At $10^5~{\rm yrs}$
before the merger, they achieve typically SNR $\sim {\rm few~}10^2$, and
achieve values as high as SNR $\sim ~10^4$ a thousand yrs before the plunge.
Since SNR is inversely proportional to the distance of the source, these
systems are also detectable in nearby galaxies-, or dwarf satellite galaxies,
with the proviso that the MBH is in the range of masses of detection, such as
Messier 32.  For a MBH mass of $M_{\rm BH}=10^6\,M_{\odot}$ and $m_{\rm
BD}=0.05\,M_{\odot}$, X-MRIs are detectable out to 50 Mpc with an SNR $=10$ 200
yrs before plunge.

A statistical treatment of the distribution of orbits in phase space which
takes into account the asymmetry of the amount of pro- and retrograde orbits on
the location of the LSO yields that every $\sim 10^5~{\rm yrs}$ one of these
objects should cross the event horizon of SgrA*. The number of X-MRIs in band
however is much larger. We have checked the results of our derivation by
simplyfying it to a single stellar mass species, the case of stellar-mass black
holes, and recover the usual results of the literature that the event rates per
year are negligible in our GC, of $\sim 10^{-6}$, even if the mass of the
stellar-mass black holes is set to $40\,M_{\odot}$.

These potential sources evolve extremely slowly, as compared to stellar-mass
black hole EMRIs.  By analysing the line density function in phase space, we
derive that there are about $15$ X-MRIs at low frequencies with high
eccentricities and associated SNRs of $\simeq ~{\rm a~few~} 100$, and about
$5$, at higher frequencies, i.e. at very high SNRs (from ${\rm a~few~} 100$ up
to $2\times 10^4$), in circular-, or almost circular orbits. These numbers can
be enhanced by a multiplyin factor of a few depending on the eccentricity of
the sources when they form.

The higher the SNR, the faster X-MRIs evolve in frequency.  However, even if it
is less likely that LISA will observe an X-MRI of SNR $20,000$ as compared to
one of a few $100$, the very loud systems live in band for as much as a few
thousand years.  A SNR of a few $100$ could already be problematic in the
detection of a binary of SMBHs, and an X-MRI with SNR of a few $1000$ could
bury the signal. Moreover, since X-MRIs can be detected for very long periodes
of time with a high eccentricity, this will be a challenge from the point of
view of data analysis, as compared to regular EMRIs. The values we have adopted
to derive these results are conservative, so that one should take into account
that X-MRIs might pose a problem if not in the detection of MBH binaries, then
in their parameter extraction.

Also, we are artificially decreasing the event rate because we are limited in
our analytical approach to pure power-laws. In numerical simulations the
power-law decreases as one approaches the innermost radii. By populating this
region with more stellar-mass black holes, we are artificially increasing
$T_{\rm rlx}$, and hence decreasing the event rate, as it can be seen in
Eq.~(\ref{eq.EventRate}). A more realistic approach should lead to an enhanced
event rate, and therefore to more sources in band.

X-MRIs are interesting because backreaction depends on the mass ratio, which
means that at $q\sim 10^8$, these systems are closer to a geodesic than EMRIs
formed with larger $q$. This means that approximations in the calculation of
the orbit are closer to the actual inspiral, and hence easier to model.  A
consequence is that it should not be difficult to separate X-MRIs signals from
(potentially) weaker ones, such as binaries of MBHs. Because they can reach
very large SNRs, and evolve very slowly in frequency, the parameter extraction
can be done in detail. {Contrary to EMRIs, X-MRIs can be regarded as
monochromatic sources for space-borne detectors. To map spacetime around
supermassive black holes in a fashion similar to EMRIs, a certain amount of
different X-MRIs would be required with different parameters such as semi-major
axes and inclinations.  A detailed parameter study such as the role of phase
accuracy in the waveform that can be achieved thanks to the high SNR is out of
the scope of this paper, and will be presented in a separate study.}

\section*{Acknowledgments}

I thank Marc Freitag and Xian Chen for
discussions, and Douglas Heggie, Pavel Kroupa, Enrico Ramiro Ru{\'i}z and
Rainer Sch{\"o}del for their kindness in replying to e-mails with questions, as
well as Sylvia Zhu, Matt Benacquista and Massimo Dotti for comments on the
manuscript. Marc also kindly helped me with the plotting programme supermongo.
I am particularly thankful with Marta Masini for her extraordinary support,
crucial during the finishing of the article. I acknowledge support from the
Ram{\'o}n y Cajal Programme of the Ministry of Economy, Industry and
Competitiveness of Spain, as well as the COST Action GWverse CA16104. This work
has been supported by the National Key R\&D Program of China (2016YFA0400702)
and the National Science Foundation of China (11721303).


\begin{thebibliography}{52}%
\makeatletter
\providecommand \@ifxundefined [1]{%
 \@ifx{#1\undefined}
}%
\providecommand \@ifnum [1]{%
 \ifnum #1\expandafter \@firstoftwo
 \else \expandafter \@secondoftwo
 \fi
}%
\providecommand \@ifx [1]{%
 \ifx #1\expandafter \@firstoftwo
 \else \expandafter \@secondoftwo
 \fi
}%
\providecommand \natexlab [1]{#1}%
\providecommand \enquote  [1]{``#1''}%
\providecommand \bibnamefont  [1]{#1}%
\providecommand \bibfnamefont [1]{#1}%
\providecommand \citenamefont [1]{#1}%
\providecommand \href@noop [0]{\@secondoftwo}%
\providecommand \href [0]{\begingroup \@sanitize@url \@href}%
\providecommand \@href[1]{\@@startlink{#1}\@@href}%
\providecommand \@@href[1]{\endgroup#1\@@endlink}%
\providecommand \@sanitize@url [0]{\catcode `\\12\catcode `\$12\catcode
  `\&12\catcode `\#12\catcode `\^12\catcode `\_12\catcode `\%12\relax}%
\providecommand \@@startlink[1]{}%
\providecommand \@@endlink[0]{}%
\providecommand \url  [0]{\begingroup\@sanitize@url \@url }%
\providecommand \@url [1]{\endgroup\@href {#1}{\urlprefix }}%
\providecommand \urlprefix  [0]{URL }%
\providecommand \Eprint [0]{\href }%
\providecommand \doibase [0]{http://dx.doi.org/}%
\providecommand \selectlanguage [0]{\@gobble}%
\providecommand \bibinfo  [0]{\@secondoftwo}%
\providecommand \bibfield  [0]{\@secondoftwo}%
\providecommand \translation [1]{[#1]}%
\providecommand \BibitemOpen [0]{}%
\providecommand \bibitemStop [0]{}%
\providecommand \bibitemNoStop [0]{.\EOS\space}%
\providecommand \EOS [0]{\spacefactor3000\relax}%
\providecommand \BibitemShut  [1]{\csname bibitem#1\endcsname}%
\let\auto@bib@innerbib\@empty
\bibitem [{\citenamefont {{Kormendy}}\ and\ \citenamefont
  {{Ho}}(2013)}]{KormendyHo2013}%
  \BibitemOpen
  \bibfield  {author} {\bibinfo {author} {\bibfnamefont {J.}~\bibnamefont
  {{Kormendy}}}\ and\ \bibinfo {author} {\bibfnamefont {L.~C.}\ \bibnamefont
  {{Ho}}},\ }\href {\doibase 10.1146/annurev-astro-082708-101811} {\bibfield
  {journal} {\bibinfo  {journal} {ARA\&A}\ }\textbf {\bibinfo {volume} {51}},\
  \bibinfo {pages} {511} (\bibinfo {year} {2013})},\ \Eprint
  {http://arxiv.org/abs/1304.7762} {arXiv:1304.7762 [astro-ph.CO]} \BibitemShut
  {NoStop}%
\bibitem [{\citenamefont {{Mezcua}}(2017)}]{Mezcua2017}%
  \BibitemOpen
  \bibfield  {author} {\bibinfo {author} {\bibfnamefont {M.}~\bibnamefont
  {{Mezcua}}},\ }\href {\doibase 10.1142/S021827181730021X} {\bibfield
  {journal} {\bibinfo  {journal} {International Journal of Modern Physics D}\
  }\textbf {\bibinfo {volume} {26}},\ \bibinfo {eid} {1730021} (\bibinfo {year}
  {2017})},\ \Eprint {http://arxiv.org/abs/1705.09667} {arXiv:1705.09667}
  \BibitemShut {NoStop}%
\bibitem [{\citenamefont {{L{\"u}tzgendorf}}\ \emph {et~al.}()\citenamefont
  {{L{\"u}tzgendorf}}, \citenamefont {{Kissler-Patig}}, \citenamefont
  {{Neumayer}}, \citenamefont {{Baumgardt}}, \citenamefont {{Noyola}},
  \citenamefont {{de Zeeuw}}, \citenamefont {{Gebhardt}}, \citenamefont
  {{Jalali}},\ and\ \citenamefont {{Feldmeier}}}]{LuetzgendorfEtAl2013}%
  \BibitemOpen
  \bibfield  {author} {\bibinfo {author} {\bibfnamefont {N.}~\bibnamefont
  {{L{\"u}tzgendorf}}}, \bibinfo {author} {\bibfnamefont {M.}~\bibnamefont
  {{Kissler-Patig}}}, \bibinfo {author} {\bibfnamefont {N.}~\bibnamefont
  {{Neumayer}}}, \bibinfo {author} {\bibfnamefont {H.}~\bibnamefont
  {{Baumgardt}}}, \bibinfo {author} {\bibfnamefont {E.}~\bibnamefont
  {{Noyola}}}, \bibinfo {author} {\bibfnamefont {P.~T.}\ \bibnamefont {{de
  Zeeuw}}}, \bibinfo {author} {\bibfnamefont {K.}~\bibnamefont {{Gebhardt}}},
  \bibinfo {author} {\bibfnamefont {B.}~\bibnamefont {{Jalali}}}, \ and\
  \bibinfo {author} {\bibfnamefont {A.}~\bibnamefont {{Feldmeier}}},\
  }\href@noop {} {\ }\BibitemShut {NoStop}%
\bibitem [{\citenamefont {{Amaro-Seoane}}\ \emph {et~al.}(2017)\citenamefont
  {{Amaro-Seoane}}, \citenamefont {{Audley}}, \citenamefont {{Babak}},
  \citenamefont {{Baker}}, \citenamefont {{Barausse}}, \citenamefont
  {{Bender}}, \citenamefont {{Berti}}, \citenamefont {{Binetruy}},
  \citenamefont {{Born}}, \citenamefont {{Bortoluzzi}}, \citenamefont {{Camp}},
  \citenamefont {{Caprini}}, \citenamefont {{Cardoso}}, \citenamefont
  {{Colpi}}, \citenamefont {{Conklin}}, \citenamefont {{Cornish}},
  \citenamefont {{Cutler}}, \citenamefont {{Danzmann}}, \citenamefont
  {{Dolesi}}, \citenamefont {{Ferraioli}}, \citenamefont {{Ferroni}},
  \citenamefont {{Fitzsimons}}, \citenamefont {{Gair}}, \citenamefont {{Gesa
  Bote}}, \citenamefont {{Giardini}}, \citenamefont {{Gibert}}, \citenamefont
  {{Grimani}}, \citenamefont {{Halloin}}, \citenamefont {{Heinzel}},
  \citenamefont {{Hertog}}, \citenamefont {{Hewitson}}, \citenamefont
  {{Holley-Bockelmann}}, \citenamefont {{Hollington}}, \citenamefont
  {{Hueller}}, \citenamefont {{Inchauspe}}, \citenamefont {{Jetzer}},
  \citenamefont {{Karnesis}}, \citenamefont {{Killow}}, \citenamefont
  {{Klein}}, \citenamefont {{Klipstein}}, \citenamefont {{Korsakova}},
  \citenamefont {{Larson}}, \citenamefont {{Livas}}, \citenamefont {{Lloro}},
  \citenamefont {{Man}}, \citenamefont {{Mance}}, \citenamefont {{Martino}},
  \citenamefont {{Mateos}}, \citenamefont {{McKenzie}}, \citenamefont
  {{McWilliams}}, \citenamefont {{Miller}}, \citenamefont {{Mueller}},
  \citenamefont {{Nardini}}, \citenamefont {{Nelemans}}, \citenamefont
  {{Nofrarias}}, \citenamefont {{Petiteau}}, \citenamefont {{Pivato}},
  \citenamefont {{Plagnol}}, \citenamefont {{Porter}}, \citenamefont
  {{Reiche}}, \citenamefont {{Robertson}}, \citenamefont {{Robertson}},
  \citenamefont {{Rossi}}, \citenamefont {{Russano}}, \citenamefont {{Schutz}},
  \citenamefont {{Sesana}}, \citenamefont {{Shoemaker}}, \citenamefont
  {{Slutsky}}, \citenamefont {{Sopuerta}}, \citenamefont {{Sumner}},
  \citenamefont {{Tamanini}}, \citenamefont {{Thorpe}}, \citenamefont
  {{Troebs}}, \citenamefont {{Vallisneri}}, \citenamefont {{Vecchio}},
  \citenamefont {{Vetrugno}}, \citenamefont {{Vitale}}, \citenamefont
  {{Volonteri}}, \citenamefont {{Wanner}}, \citenamefont {{Ward}},
  \citenamefont {{Wass}}, \citenamefont {{Weber}}, \citenamefont {{Ziemer}},\
  and\ \citenamefont {{Zweifel}}}]{Amaro-SeoaneEtAl2017}%
  \BibitemOpen
  \bibfield  {author} {\bibinfo {author} {\bibfnamefont {P.}~\bibnamefont
  {{Amaro-Seoane}}}, \bibinfo {author} {\bibfnamefont {H.}~\bibnamefont
  {{Audley}}}, \bibinfo {author} {\bibfnamefont {S.}~\bibnamefont {{Babak}}},
  \bibinfo {author} {\bibfnamefont {J.}~\bibnamefont {{Baker}}}, \bibinfo
  {author} {\bibfnamefont {E.}~\bibnamefont {{Barausse}}}, \bibinfo {author}
  {\bibfnamefont {P.}~\bibnamefont {{Bender}}}, \bibinfo {author}
  {\bibfnamefont {E.}~\bibnamefont {{Berti}}}, \bibinfo {author} {\bibfnamefont
  {P.}~\bibnamefont {{Binetruy}}}, \bibinfo {author} {\bibfnamefont
  {M.}~\bibnamefont {{Born}}}, \bibinfo {author} {\bibfnamefont
  {D.}~\bibnamefont {{Bortoluzzi}}}, \bibinfo {author} {\bibfnamefont
  {J.}~\bibnamefont {{Camp}}}, \bibinfo {author} {\bibfnamefont
  {C.}~\bibnamefont {{Caprini}}}, \bibinfo {author} {\bibfnamefont
  {V.}~\bibnamefont {{Cardoso}}}, \bibinfo {author} {\bibfnamefont
  {M.}~\bibnamefont {{Colpi}}}, \bibinfo {author} {\bibfnamefont
  {J.}~\bibnamefont {{Conklin}}}, \bibinfo {author} {\bibfnamefont
  {N.}~\bibnamefont {{Cornish}}}, \bibinfo {author} {\bibfnamefont
  {C.}~\bibnamefont {{Cutler}}}, \bibinfo {author} {\bibfnamefont
  {K.}~\bibnamefont {{Danzmann}}}, \bibinfo {author} {\bibfnamefont
  {R.}~\bibnamefont {{Dolesi}}}, \bibinfo {author} {\bibfnamefont
  {L.}~\bibnamefont {{Ferraioli}}}, \bibinfo {author} {\bibfnamefont
  {V.}~\bibnamefont {{Ferroni}}}, \bibinfo {author} {\bibfnamefont
  {E.}~\bibnamefont {{Fitzsimons}}}, \bibinfo {author} {\bibfnamefont
  {J.}~\bibnamefont {{Gair}}}, \bibinfo {author} {\bibfnamefont
  {L.}~\bibnamefont {{Gesa Bote}}}, \bibinfo {author} {\bibfnamefont
  {D.}~\bibnamefont {{Giardini}}}, \bibinfo {author} {\bibfnamefont
  {F.}~\bibnamefont {{Gibert}}}, \bibinfo {author} {\bibfnamefont
  {C.}~\bibnamefont {{Grimani}}}, \bibinfo {author} {\bibfnamefont
  {H.}~\bibnamefont {{Halloin}}}, \bibinfo {author} {\bibfnamefont
  {G.}~\bibnamefont {{Heinzel}}}, \bibinfo {author} {\bibfnamefont
  {T.}~\bibnamefont {{Hertog}}}, \bibinfo {author} {\bibfnamefont
  {M.}~\bibnamefont {{Hewitson}}}, \bibinfo {author} {\bibfnamefont
  {K.}~\bibnamefont {{Holley-Bockelmann}}}, \bibinfo {author} {\bibfnamefont
  {D.}~\bibnamefont {{Hollington}}}, \bibinfo {author} {\bibfnamefont
  {M.}~\bibnamefont {{Hueller}}}, \bibinfo {author} {\bibfnamefont
  {H.}~\bibnamefont {{Inchauspe}}}, \bibinfo {author} {\bibfnamefont
  {P.}~\bibnamefont {{Jetzer}}}, \bibinfo {author} {\bibfnamefont
  {N.}~\bibnamefont {{Karnesis}}}, \bibinfo {author} {\bibfnamefont
  {C.}~\bibnamefont {{Killow}}}, \bibinfo {author} {\bibfnamefont
  {A.}~\bibnamefont {{Klein}}}, \bibinfo {author} {\bibfnamefont
  {B.}~\bibnamefont {{Klipstein}}}, \bibinfo {author} {\bibfnamefont
  {N.}~\bibnamefont {{Korsakova}}}, \bibinfo {author} {\bibfnamefont {S.~L.}\
  \bibnamefont {{Larson}}}, \bibinfo {author} {\bibfnamefont {J.}~\bibnamefont
  {{Livas}}}, \bibinfo {author} {\bibfnamefont {I.}~\bibnamefont {{Lloro}}},
  \bibinfo {author} {\bibfnamefont {N.}~\bibnamefont {{Man}}}, \bibinfo
  {author} {\bibfnamefont {D.}~\bibnamefont {{Mance}}}, \bibinfo {author}
  {\bibfnamefont {J.}~\bibnamefont {{Martino}}}, \bibinfo {author}
  {\bibfnamefont {I.}~\bibnamefont {{Mateos}}}, \bibinfo {author}
  {\bibfnamefont {K.}~\bibnamefont {{McKenzie}}}, \bibinfo {author}
  {\bibfnamefont {S.~T.}\ \bibnamefont {{McWilliams}}}, \bibinfo {author}
  {\bibfnamefont {C.}~\bibnamefont {{Miller}}}, \bibinfo {author}
  {\bibfnamefont {G.}~\bibnamefont {{Mueller}}}, \bibinfo {author}
  {\bibfnamefont {G.}~\bibnamefont {{Nardini}}}, \bibinfo {author}
  {\bibfnamefont {G.}~\bibnamefont {{Nelemans}}}, \bibinfo {author}
  {\bibfnamefont {M.}~\bibnamefont {{Nofrarias}}}, \bibinfo {author}
  {\bibfnamefont {A.}~\bibnamefont {{Petiteau}}}, \bibinfo {author}
  {\bibfnamefont {P.}~\bibnamefont {{Pivato}}}, \bibinfo {author}
  {\bibfnamefont {E.}~\bibnamefont {{Plagnol}}}, \bibinfo {author}
  {\bibfnamefont {E.}~\bibnamefont {{Porter}}}, \bibinfo {author}
  {\bibfnamefont {J.}~\bibnamefont {{Reiche}}}, \bibinfo {author}
  {\bibfnamefont {D.}~\bibnamefont {{Robertson}}}, \bibinfo {author}
  {\bibfnamefont {N.}~\bibnamefont {{Robertson}}}, \bibinfo {author}
  {\bibfnamefont {E.}~\bibnamefont {{Rossi}}}, \bibinfo {author} {\bibfnamefont
  {G.}~\bibnamefont {{Russano}}}, \bibinfo {author} {\bibfnamefont
  {B.}~\bibnamefont {{Schutz}}}, \bibinfo {author} {\bibfnamefont
  {A.}~\bibnamefont {{Sesana}}}, \bibinfo {author} {\bibfnamefont
  {D.}~\bibnamefont {{Shoemaker}}}, \bibinfo {author} {\bibfnamefont
  {J.}~\bibnamefont {{Slutsky}}}, \bibinfo {author} {\bibfnamefont {C.~F.}\
  \bibnamefont {{Sopuerta}}}, \bibinfo {author} {\bibfnamefont
  {T.}~\bibnamefont {{Sumner}}}, \bibinfo {author} {\bibfnamefont
  {N.}~\bibnamefont {{Tamanini}}}, \bibinfo {author} {\bibfnamefont
  {I.}~\bibnamefont {{Thorpe}}}, \bibinfo {author} {\bibfnamefont
  {M.}~\bibnamefont {{Troebs}}}, \bibinfo {author} {\bibfnamefont
  {M.}~\bibnamefont {{Vallisneri}}}, \bibinfo {author} {\bibfnamefont
  {A.}~\bibnamefont {{Vecchio}}}, \bibinfo {author} {\bibfnamefont
  {D.}~\bibnamefont {{Vetrugno}}}, \bibinfo {author} {\bibfnamefont
  {S.}~\bibnamefont {{Vitale}}}, \bibinfo {author} {\bibfnamefont
  {M.}~\bibnamefont {{Volonteri}}}, \bibinfo {author} {\bibfnamefont
  {G.}~\bibnamefont {{Wanner}}}, \bibinfo {author} {\bibfnamefont
  {H.}~\bibnamefont {{Ward}}}, \bibinfo {author} {\bibfnamefont
  {P.}~\bibnamefont {{Wass}}}, \bibinfo {author} {\bibfnamefont
  {W.}~\bibnamefont {{Weber}}}, \bibinfo {author} {\bibfnamefont
  {J.}~\bibnamefont {{Ziemer}}}, \ and\ \bibinfo {author} {\bibfnamefont
  {P.}~\bibnamefont {{Zweifel}}},\ }\href@noop {} {\bibfield  {journal}
  {\bibinfo  {journal} {ArXiv e-prints}\ } (\bibinfo {year} {2017})},\ \Eprint
  {http://arxiv.org/abs/1702.00786} {arXiv:1702.00786 [astro-ph.IM]}
  \BibitemShut {NoStop}%
\bibitem [{\citenamefont
  {{Amaro-Seoane}}(2018{\natexlab{a}})}]{Amaro-SeoaneLRR2012}%
  \BibitemOpen
  \bibfield  {author} {\bibinfo {author} {\bibfnamefont {P.}~\bibnamefont
  {{Amaro-Seoane}}},\ }\href {\doibase 10.1007/s41114-018-0013-8} {\bibfield
  {journal} {\bibinfo  {journal} {Living Reviews in Relativity}\ }\textbf
  {\bibinfo {volume} {21}},\ \bibinfo {eid} {4} (\bibinfo {year}
  {2018}{\natexlab{a}})},\ \Eprint {http://arxiv.org/abs/1205.5240}
  {arXiv:1205.5240} \BibitemShut {NoStop}%
\bibitem [{\citenamefont {{Amaro-Seoane}}\ \emph {et~al.}(2015)\citenamefont
  {{Amaro-Seoane}}, \citenamefont {{Gair}}, \citenamefont {{Pound}},
  \citenamefont {{Hughes}},\ and\ \citenamefont
  {{Sopuerta}}}]{Amaro-SeoaneGairPoundHughesSopuerta2015}%
  \BibitemOpen
  \bibfield  {author} {\bibinfo {author} {\bibfnamefont {P.}~\bibnamefont
  {{Amaro-Seoane}}}, \bibinfo {author} {\bibfnamefont {J.~R.}\ \bibnamefont
  {{Gair}}}, \bibinfo {author} {\bibfnamefont {A.}~\bibnamefont {{Pound}}},
  \bibinfo {author} {\bibfnamefont {S.~A.}\ \bibnamefont {{Hughes}}}, \ and\
  \bibinfo {author} {\bibfnamefont {C.~F.}\ \bibnamefont {{Sopuerta}}},\ }\href
  {\doibase 10.1088/1742-6596/610/1/012002} {\bibfield  {journal} {\bibinfo
  {journal} {Journal of Physics Conference Series}\ }\textbf {\bibinfo {volume}
  {610}},\ \bibinfo {eid} {012002} (\bibinfo {year} {2015})},\ \Eprint
  {http://arxiv.org/abs/1410.0958} {arXiv:1410.0958} \BibitemShut {NoStop}%
\bibitem [{\citenamefont
  {{Amaro-Seoane}}(2018{\natexlab{b}})}]{Amaro-Seoane2018}%
  \BibitemOpen
  \bibfield  {author} {\bibinfo {author} {\bibfnamefont {P.}~\bibnamefont
  {{Amaro-Seoane}}},\ }\href {\doibase 10.1103/PhysRevD.98.063018} {\bibfield
  {journal} {\bibinfo  {journal} {Phys.Rev.D.}\ }\textbf {\bibinfo {volume}
  {98}},\ \bibinfo {eid} {063018} (\bibinfo {year} {2018}{\natexlab{b}})},\
  \Eprint {http://arxiv.org/abs/1807.03824} {arXiv:1807.03824 [astro-ph.HE]}
  \BibitemShut {NoStop}%
\bibitem [{\citenamefont {{Hills}}(1975)}]{Hills75}%
  \BibitemOpen
  \bibfield  {author} {\bibinfo {author} {\bibfnamefont {J.~G.}\ \bibnamefont
  {{Hills}}},\ }\href@noop {} {\bibfield  {journal} {\bibinfo  {journal} {Nat}\
  }\textbf {\bibinfo {volume} {254}},\ \bibinfo {pages} {295} (\bibinfo {year}
  {1975})}\BibitemShut {NoStop}%
\bibitem [{\citenamefont {{Rees}}(1988)}]{Rees88}%
  \BibitemOpen
  \bibfield  {author} {\bibinfo {author} {\bibfnamefont {M.~J.}\ \bibnamefont
  {{Rees}}},\ }\href {\doibase 10.1038/333523a0} {\bibfield  {journal}
  {\bibinfo  {journal} {Nat}\ }\textbf {\bibinfo {volume} {333}},\ \bibinfo
  {pages} {523} (\bibinfo {year} {1988})}\BibitemShut {NoStop}%
\bibitem [{\citenamefont {{Shapiro}}\ and\ \citenamefont
  {{Teukolsky}}(1983)}]{ST83}%
  \BibitemOpen
  \bibfield  {author} {\bibinfo {author} {\bibfnamefont {S.~L.}\ \bibnamefont
  {{Shapiro}}}\ and\ \bibinfo {author} {\bibfnamefont {S.~A.}\ \bibnamefont
  {{Teukolsky}}},\ }\href@noop {} {\emph {\bibinfo {title} {Black holes, white
  dwarfs, and neutron stars: The physics of compact objects}}}\ (\bibinfo
  {publisher} {Wiley-Interscience},\ \bibinfo {year} {1983})\BibitemShut
  {NoStop}%
\bibitem [{\citenamefont {{Chandrasekhar}}(1942)}]{Chandra42}%
  \BibitemOpen
  \bibfield  {author} {\bibinfo {author} {\bibfnamefont {S.}~\bibnamefont
  {{Chandrasekhar}}},\ }\href@noop {} {\bibfield  {journal} {\bibinfo
  {journal} {Physical Sciences Data}\ } (\bibinfo {year} {1942})}\BibitemShut
  {NoStop}%
\bibitem [{\citenamefont {{Schaller}}\ \emph {et~al.}(1992)\citenamefont
  {{Schaller}}, \citenamefont {{Schaerer}}, \citenamefont {{Meynet}},\ and\
  \citenamefont {{Maeder}}}]{SSMM92}%
  \BibitemOpen
  \bibfield  {author} {\bibinfo {author} {\bibfnamefont {G.}~\bibnamefont
  {{Schaller}}}, \bibinfo {author} {\bibfnamefont {D.}~\bibnamefont
  {{Schaerer}}}, \bibinfo {author} {\bibfnamefont {G.}~\bibnamefont
  {{Meynet}}}, \ and\ \bibinfo {author} {\bibfnamefont {A.}~\bibnamefont
  {{Maeder}}},\ }\href@noop {} {\bibfield  {journal} {\bibinfo  {journal}
  {A\&AS}\ }\textbf {\bibinfo {volume} {96}},\ \bibinfo {pages} {269} (\bibinfo
  {year} {1992})}\BibitemShut {NoStop}%
\bibitem [{\citenamefont {{Meynet}}\ \emph {et~al.}(1994)\citenamefont
  {{Meynet}}, \citenamefont {{Maeder}}, \citenamefont {{Schaller}},
  \citenamefont {{Schaerer}},\ and\ \citenamefont {{Charbonnel}}}]{MMSSC94}%
  \BibitemOpen
  \bibfield  {author} {\bibinfo {author} {\bibfnamefont {G.}~\bibnamefont
  {{Meynet}}}, \bibinfo {author} {\bibfnamefont {A.}~\bibnamefont {{Maeder}}},
  \bibinfo {author} {\bibfnamefont {G.}~\bibnamefont {{Schaller}}}, \bibinfo
  {author} {\bibfnamefont {D.}~\bibnamefont {{Schaerer}}}, \ and\ \bibinfo
  {author} {\bibfnamefont {C.}~\bibnamefont {{Charbonnel}}},\ }\href@noop {}
  {\bibfield  {journal} {\bibinfo  {journal} {A\&AS}\ }\textbf {\bibinfo
  {volume} {103}},\ \bibinfo {pages} {97} (\bibinfo {year} {1994})}\BibitemShut
  {NoStop}%
\bibitem [{\citenamefont {{Charbonnel}}\ \emph {et~al.}(1999)\citenamefont
  {{Charbonnel}}, \citenamefont {{D\"{a}ppen}}, \citenamefont {{Schaerer}},
  \citenamefont {{Bernasconi}}, \citenamefont {{Maeder}}, \citenamefont
  {{Meynet}},\ and\ \citenamefont {{Mowlavi}}}]{CDSBMMM99}%
  \BibitemOpen
  \bibfield  {author} {\bibinfo {author} {\bibfnamefont {C.}~\bibnamefont
  {{Charbonnel}}}, \bibinfo {author} {\bibfnamefont {W.}~\bibnamefont
  {{D\"{a}ppen}}}, \bibinfo {author} {\bibfnamefont {D.}~\bibnamefont
  {{Schaerer}}}, \bibinfo {author} {\bibfnamefont {P.~A.}\ \bibnamefont
  {{Bernasconi}}}, \bibinfo {author} {\bibfnamefont {A.}~\bibnamefont
  {{Maeder}}}, \bibinfo {author} {\bibfnamefont {G.}~\bibnamefont {{Meynet}}},
  \ and\ \bibinfo {author} {\bibfnamefont {N.}~\bibnamefont {{Mowlavi}}},\
  }\href@noop {} {\bibfield  {journal} {\bibinfo  {journal} {A\&AS}\ }\textbf
  {\bibinfo {volume} {135}},\ \bibinfo {pages} {405} (\bibinfo {year}
  {1999})}\BibitemShut {NoStop}%
\bibitem [{\citenamefont {{Chabrier}}\ and\ \citenamefont
  {{Baraffe}}(2000)}]{CB00}%
  \BibitemOpen
  \bibfield  {author} {\bibinfo {author} {\bibfnamefont {G.}~\bibnamefont
  {{Chabrier}}}\ and\ \bibinfo {author} {\bibfnamefont {I.}~\bibnamefont
  {{Baraffe}}},\ }\href@noop {} {\bibfield  {journal} {\bibinfo  {journal}
  {ARA\&A}\ }\textbf {\bibinfo {volume} {38}},\ \bibinfo {pages} {337}
  (\bibinfo {year} {2000})}\BibitemShut {NoStop}%
\bibitem [{\citenamefont {{Chabrier}}\ \emph {et~al.}(2009)\citenamefont
  {{Chabrier}}, \citenamefont {{Baraffe}}, \citenamefont {{Leconte}},
  \citenamefont {{Gallardo}},\ and\ \citenamefont
  {{Barman}}}]{ChabrierEtAl2009}%
  \BibitemOpen
  \bibfield  {author} {\bibinfo {author} {\bibfnamefont {G.}~\bibnamefont
  {{Chabrier}}}, \bibinfo {author} {\bibfnamefont {I.}~\bibnamefont
  {{Baraffe}}}, \bibinfo {author} {\bibfnamefont {J.}~\bibnamefont
  {{Leconte}}}, \bibinfo {author} {\bibfnamefont {J.}~\bibnamefont
  {{Gallardo}}}, \ and\ \bibinfo {author} {\bibfnamefont {T.}~\bibnamefont
  {{Barman}}},\ }in\ \href {\doibase 10.1063/1.3099078} {\emph {\bibinfo
  {booktitle} {15th Cambridge Workshop on Cool Stars, Stellar Systems, and the
  Sun}}},\ \bibinfo {series} {American Institute of Physics Conference Series},
  Vol.\ \bibinfo {volume} {1094},\ \bibinfo {editor} {edited by\ \bibinfo
  {editor} {\bibfnamefont {E.}~\bibnamefont {{Stempels}}}}\ (\bibinfo {year}
  {2009})\ pp.\ \bibinfo {pages} {102--111},\ \Eprint
  {http://arxiv.org/abs/0810.5085} {arXiv:0810.5085} \BibitemShut {NoStop}%
\bibitem [{\citenamefont {{Freitag}}(2003)}]{Freitag03}%
  \BibitemOpen
  \bibfield  {author} {\bibinfo {author} {\bibfnamefont {M.}~\bibnamefont
  {{Freitag}}},\ }\href {\doibase 10.1086/367813} {\bibfield  {journal}
  {\bibinfo  {journal} {ApJ Lett.}\ }\textbf {\bibinfo {volume} {583}},\
  \bibinfo {pages} {L21} (\bibinfo {year} {2003})},\ \Eprint
  {http://arxiv.org/abs/arXiv:astro-ph/0211209} {arXiv:astro-ph/0211209}
  \BibitemShut {NoStop}%
\bibitem [{\citenamefont {{Barack}}\ and\ \citenamefont
  {{Cutler}}(2004)}]{BarackCutler2004}%
  \BibitemOpen
  \bibfield  {author} {\bibinfo {author} {\bibfnamefont {L.}~\bibnamefont
  {{Barack}}}\ and\ \bibinfo {author} {\bibfnamefont {C.}~\bibnamefont
  {{Cutler}}},\ }\href {\doibase 10.1103/PhysRevD.69.082005} {\bibfield
  {journal} {\bibinfo  {journal} {Phys. Rev. D}\ }\textbf {\bibinfo {volume}
  {69}},\ \bibinfo {eid} {082005} (\bibinfo {year} {2004})},\ \Eprint
  {http://arxiv.org/abs/gr-qc/0310125} {gr-qc/0310125} \BibitemShut {NoStop}%
\bibitem [{\citenamefont {{Gourgoulhon}}\ \emph {et~al.}(2019)\citenamefont
  {{Gourgoulhon}}, \citenamefont {{Le Tiec}}, \citenamefont {{Vincent}},\ and\
  \citenamefont {{Warburton}}}]{GourgoulhonEtAl2019}%
  \BibitemOpen
  \bibfield  {author} {\bibinfo {author} {\bibfnamefont {E.}~\bibnamefont
  {{Gourgoulhon}}}, \bibinfo {author} {\bibfnamefont {A.}~\bibnamefont {{Le
  Tiec}}}, \bibinfo {author} {\bibfnamefont {F.~H.}\ \bibnamefont {{Vincent}}},
  \ and\ \bibinfo {author} {\bibfnamefont {N.}~\bibnamefont {{Warburton}}},\
  }\href@noop {} {\bibfield  {journal} {\bibinfo  {journal} {arXiv e-prints}\
  ,\ \bibinfo {eid} {arXiv:1903.02049}} (\bibinfo {year} {2019})},\ \Eprint
  {http://arxiv.org/abs/1903.02049} {arXiv:1903.02049 [gr-qc]} \BibitemShut
  {NoStop}%
\bibitem [{\citenamefont {{Peebles}}(1972)}]{Peebles1972}%
  \BibitemOpen
  \bibfield  {author} {\bibinfo {author} {\bibfnamefont {P.~J.~E.}\
  \bibnamefont {{Peebles}}},\ }\href {\doibase 10.1086/151797} {\bibfield
  {journal} {\bibinfo  {journal} {ApJ}\ }\textbf {\bibinfo {volume} {178}},\
  \bibinfo {pages} {371} (\bibinfo {year} {1972})}\BibitemShut {NoStop}%
\bibitem [{\citenamefont {{Bahcall}}\ and\ \citenamefont
  {{Wolf}}(1976)}]{BW76}%
  \BibitemOpen
  \bibfield  {author} {\bibinfo {author} {\bibfnamefont {J.~N.}\ \bibnamefont
  {{Bahcall}}}\ and\ \bibinfo {author} {\bibfnamefont {R.~A.}\ \bibnamefont
  {{Wolf}}},\ }\href@noop {} {\bibfield  {journal} {\bibinfo  {journal} {ApJ}\
  }\textbf {\bibinfo {volume} {209}},\ \bibinfo {pages} {214} (\bibinfo {year}
  {1976})}\BibitemShut {NoStop}%
\bibitem [{\citenamefont {{Gurevich}}(1964)}]{Gurevich64}%
  \BibitemOpen
  \bibfield  {author} {\bibinfo {author} {\bibfnamefont {A.}~\bibnamefont
  {{Gurevich}}},\ }\href@noop {} {\bibfield  {journal} {\bibinfo  {journal}
  {Geomag. Aeronom.}\ }\textbf {\bibinfo {volume} {4}},\ \bibinfo {pages} {247}
  (\bibinfo {year} {1964})}\BibitemShut {NoStop}%
\bibitem [{\citenamefont {{Shapiro}}\ and\ \citenamefont
  {{Marchant}}(1978)}]{SM78}%
  \BibitemOpen
  \bibfield  {author} {\bibinfo {author} {\bibfnamefont {S.~L.}\ \bibnamefont
  {{Shapiro}}}\ and\ \bibinfo {author} {\bibfnamefont {A.~B.}\ \bibnamefont
  {{Marchant}}},\ }\href@noop {} {\bibfield  {journal} {\bibinfo  {journal}
  {ApJ}\ }\textbf {\bibinfo {volume} {225}},\ \bibinfo {pages} {603} (\bibinfo
  {year} {1978})}\BibitemShut {NoStop}%
\bibitem [{\citenamefont {{Marchant}}\ and\ \citenamefont
  {{Shapiro}}(1979)}]{MS79}%
  \BibitemOpen
  \bibfield  {author} {\bibinfo {author} {\bibfnamefont {A.~B.}\ \bibnamefont
  {{Marchant}}}\ and\ \bibinfo {author} {\bibfnamefont {S.~L.}\ \bibnamefont
  {{Shapiro}}},\ }\href@noop {} {\bibfield  {journal} {\bibinfo  {journal}
  {ApJ}\ }\textbf {\bibinfo {volume} {234}},\ \bibinfo {pages} {317} (\bibinfo
  {year} {1979})}\BibitemShut {NoStop}%
\bibitem [{\citenamefont {{Marchant}}\ and\ \citenamefont
  {{Shapiro}}(1980)}]{MS80}%
  \BibitemOpen
  \bibfield  {author} {\bibinfo {author} {\bibfnamefont {A.~B.}\ \bibnamefont
  {{Marchant}}}\ and\ \bibinfo {author} {\bibfnamefont {S.~L.}\ \bibnamefont
  {{Shapiro}}},\ }\href@noop {} {\bibfield  {journal} {\bibinfo  {journal}
  {ApJ}\ }\textbf {\bibinfo {volume} {239}},\ \bibinfo {pages} {685} (\bibinfo
  {year} {1980})}\BibitemShut {NoStop}%
\bibitem [{\citenamefont {{Shapiro}}\ and\ \citenamefont
  {{Teukolsky}}(1985)}]{ST85}%
  \BibitemOpen
  \bibfield  {author} {\bibinfo {author} {\bibfnamefont {S.~L.}\ \bibnamefont
  {{Shapiro}}}\ and\ \bibinfo {author} {\bibfnamefont {S.~A.}\ \bibnamefont
  {{Teukolsky}}},\ }\href@noop {} {\bibfield  {journal} {\bibinfo  {journal}
  {ApJ Lett.}\ }\textbf {\bibinfo {volume} {292}},\ \bibinfo {pages} {41}
  (\bibinfo {year} {1985})}\BibitemShut {NoStop}%
\bibitem [{\citenamefont {{Freitag}}\ and\ \citenamefont
  {{Benz}}(2001)}]{FB01a}%
  \BibitemOpen
  \bibfield  {author} {\bibinfo {author} {\bibfnamefont {M.}~\bibnamefont
  {{Freitag}}}\ and\ \bibinfo {author} {\bibfnamefont {W.}~\bibnamefont
  {{Benz}}},\ }\href@noop {} {\bibfield  {journal} {\bibinfo  {journal} {A\&A}\
  }\textbf {\bibinfo {volume} {375}},\ \bibinfo {pages} {711} (\bibinfo {year}
  {2001})}\BibitemShut {NoStop}%
\bibitem [{\citenamefont {{Amaro-Seoane}}\ \emph {et~al.}(2004)\citenamefont
  {{Amaro-Seoane}}, \citenamefont {{Freitag}},\ and\ \citenamefont
  {{Spurzem}}}]{ASEtAl04}%
  \BibitemOpen
  \bibfield  {author} {\bibinfo {author} {\bibfnamefont {P.}~\bibnamefont
  {{Amaro-Seoane}}}, \bibinfo {author} {\bibfnamefont {M.}~\bibnamefont
  {{Freitag}}}, \ and\ \bibinfo {author} {\bibfnamefont {R.}~\bibnamefont
  {{Spurzem}}},\ }\href@noop {} {\bibfield  {journal} {\bibinfo  {journal}
  {MNRAS}\ } (\bibinfo {year} {2004})},\ \Eprint
  {http://arxiv.org/abs/astro-ph/0401163} {astro-ph/0401163} \BibitemShut
  {NoStop}%
\bibitem [{\citenamefont {Preto}\ \emph {et~al.}(2004)\citenamefont {Preto},
  \citenamefont {Merritt},\ and\ \citenamefont
  {Spurzem}}]{PretoMerrittSpurzem04}%
  \BibitemOpen
  \bibfield  {author} {\bibinfo {author} {\bibfnamefont {M.}~\bibnamefont
  {Preto}}, \bibinfo {author} {\bibfnamefont {D.}~\bibnamefont {Merritt}}, \
  and\ \bibinfo {author} {\bibfnamefont {R.}~\bibnamefont {Spurzem}},\ }\href
  {\doibase 10.1086/425139} {\bibfield  {journal} {\bibinfo  {journal} {ApJ
  Lett.}\ }\textbf {\bibinfo {volume} {613}},\ \bibinfo {pages} {L109}
  (\bibinfo {year} {2004})}\BibitemShut {NoStop}%
\bibitem [{\citenamefont {{Baumgardt}}\ \emph {et~al.}(2018)\citenamefont
  {{Baumgardt}}, \citenamefont {{Amaro-Seoane}},\ and\ \citenamefont
  {{Sch{\"o}del}}}]{BaumgardtEtAl2018}%
  \BibitemOpen
  \bibfield  {author} {\bibinfo {author} {\bibfnamefont {H.}~\bibnamefont
  {{Baumgardt}}}, \bibinfo {author} {\bibfnamefont {P.}~\bibnamefont
  {{Amaro-Seoane}}}, \ and\ \bibinfo {author} {\bibfnamefont {R.}~\bibnamefont
  {{Sch{\"o}del}}},\ }\href {\doibase 10.1051/0004-6361/201730462} {\bibfield
  {journal} {\bibinfo  {journal} {A\&A}\ }\textbf {\bibinfo {volume} {609}},\
  \bibinfo {eid} {A28} (\bibinfo {year} {2018})},\ \Eprint
  {http://arxiv.org/abs/1701.03818} {arXiv:1701.03818} \BibitemShut {NoStop}%
\bibitem [{\citenamefont {{Gallego-Cano}}\ \emph {et~al.}(2018)\citenamefont
  {{Gallego-Cano}}, \citenamefont {{Sch{\"o}del}}, \citenamefont {{Dong}},
  \citenamefont {{Nogueras-Lara}}, \citenamefont {{Gallego-Calvente}},
  \citenamefont {{Amaro-Seoane}},\ and\ \citenamefont
  {{Baumgardt}}}]{Gallego-CanoEtAl2018}%
  \BibitemOpen
  \bibfield  {author} {\bibinfo {author} {\bibfnamefont {E.}~\bibnamefont
  {{Gallego-Cano}}}, \bibinfo {author} {\bibfnamefont {R.}~\bibnamefont
  {{Sch{\"o}del}}}, \bibinfo {author} {\bibfnamefont {H.}~\bibnamefont
  {{Dong}}}, \bibinfo {author} {\bibfnamefont {F.}~\bibnamefont
  {{Nogueras-Lara}}}, \bibinfo {author} {\bibfnamefont {A.~T.}\ \bibnamefont
  {{Gallego-Calvente}}}, \bibinfo {author} {\bibfnamefont {P.}~\bibnamefont
  {{Amaro-Seoane}}}, \ and\ \bibinfo {author} {\bibfnamefont {H.}~\bibnamefont
  {{Baumgardt}}},\ }\href {\doibase 10.1051/0004-6361/201730451} {\bibfield
  {journal} {\bibinfo  {journal} {A\&A}\ }\textbf {\bibinfo {volume} {609}},\
  \bibinfo {eid} {A26} (\bibinfo {year} {2018})},\ \Eprint
  {http://arxiv.org/abs/1701.03816} {arXiv:1701.03816} \BibitemShut {NoStop}%
\bibitem [{\citenamefont {{Sch{\"o}del}}\ \emph {et~al.}(2018)\citenamefont
  {{Sch{\"o}del}}, \citenamefont {{Gallego-Cano}}, \citenamefont {{Dong}},
  \citenamefont {{Nogueras-Lara}}, \citenamefont {{Gallego-Calvente}},
  \citenamefont {{Amaro-Seoane}},\ and\ \citenamefont
  {{Baumgardt}}}]{SchoedelEtAl2018}%
  \BibitemOpen
  \bibfield  {author} {\bibinfo {author} {\bibfnamefont {R.}~\bibnamefont
  {{Sch{\"o}del}}}, \bibinfo {author} {\bibfnamefont {E.}~\bibnamefont
  {{Gallego-Cano}}}, \bibinfo {author} {\bibfnamefont {H.}~\bibnamefont
  {{Dong}}}, \bibinfo {author} {\bibfnamefont {F.}~\bibnamefont
  {{Nogueras-Lara}}}, \bibinfo {author} {\bibfnamefont {A.~T.}\ \bibnamefont
  {{Gallego-Calvente}}}, \bibinfo {author} {\bibfnamefont {P.}~\bibnamefont
  {{Amaro-Seoane}}}, \ and\ \bibinfo {author} {\bibfnamefont {H.}~\bibnamefont
  {{Baumgardt}}},\ }\href {\doibase 10.1051/0004-6361/201730452} {\bibfield
  {journal} {\bibinfo  {journal} {A\&A}\ }\textbf {\bibinfo {volume} {609}},\
  \bibinfo {eid} {A27} (\bibinfo {year} {2018})},\ \Eprint
  {http://arxiv.org/abs/1701.03817} {arXiv:1701.03817} \BibitemShut {NoStop}%
\bibitem [{\citenamefont {{Kroupa}}\ \emph {et~al.}(2013)\citenamefont
  {{Kroupa}}, \citenamefont {{Weidner}}, \citenamefont {{Pflamm-Altenburg}},
  \citenamefont {{Thies}}, \citenamefont {{Dabringhausen}}, \citenamefont
  {{Marks}},\ and\ \citenamefont {{Maschberger}}}]{KroupaEtAl2013}%
  \BibitemOpen
  \bibfield  {author} {\bibinfo {author} {\bibfnamefont {P.}~\bibnamefont
  {{Kroupa}}}, \bibinfo {author} {\bibfnamefont {C.}~\bibnamefont {{Weidner}}},
  \bibinfo {author} {\bibfnamefont {J.}~\bibnamefont {{Pflamm-Altenburg}}},
  \bibinfo {author} {\bibfnamefont {I.}~\bibnamefont {{Thies}}}, \bibinfo
  {author} {\bibfnamefont {J.}~\bibnamefont {{Dabringhausen}}}, \bibinfo
  {author} {\bibfnamefont {M.}~\bibnamefont {{Marks}}}, \ and\ \bibinfo
  {author} {\bibfnamefont {T.}~\bibnamefont {{Maschberger}}},\ }\enquote
  {\bibinfo {title} {{The Stellar and Sub-Stellar Initial Mass Function of
  Simple and Composite Populations}},}\ in\ \href {\doibase
  10.1007/978-94-007-5612-0_4} {\emph {\bibinfo {booktitle} {Planets, Stars and
  Stellar Systems.~Volume 5: Galactic Structure and Stellar Populations}}},\
  \bibinfo {editor} {edited by\ \bibinfo {editor} {\bibfnamefont {T.~D.}\
  \bibnamefont {{Oswalt}}}\ and\ \bibinfo {editor} {\bibfnamefont
  {G.}~\bibnamefont {{Gilmore}}}}\ (\bibinfo {year} {2013})\ p.\ \bibinfo
  {pages} {115}\BibitemShut {NoStop}%
\bibitem [{\citenamefont {{Wegg}}\ \emph {et~al.}(2017)\citenamefont {{Wegg}},
  \citenamefont {{Gerhard}},\ and\ \citenamefont {{Portail}}}]{WeggEtAl2017}%
  \BibitemOpen
  \bibfield  {author} {\bibinfo {author} {\bibfnamefont {C.}~\bibnamefont
  {{Wegg}}}, \bibinfo {author} {\bibfnamefont {O.}~\bibnamefont {{Gerhard}}}, \
  and\ \bibinfo {author} {\bibfnamefont {M.}~\bibnamefont {{Portail}}},\ }\href
  {\doibase 10.3847/2041-8213/aa794e} {\bibfield  {journal} {\bibinfo
  {journal} {ApJ Lett.}\ }\textbf {\bibinfo {volume} {843}},\ \bibinfo {eid}
  {L5} (\bibinfo {year} {2017})},\ \Eprint {http://arxiv.org/abs/1706.04193}
  {arXiv:1706.04193} \BibitemShut {NoStop}%
\bibitem [{\citenamefont {{Bartko}}\ \emph {et~al.}(2010)\citenamefont
  {{Bartko}}, \citenamefont {{Martins}}, \citenamefont {{Trippe}},
  \citenamefont {{Fritz}}, \citenamefont {{Genzel}}, \citenamefont {{Ott}},
  \citenamefont {{Eisenhauer}}, \citenamefont {{Gillessen}}, \citenamefont
  {{Paumard}}, \citenamefont {{Alexander}}, \citenamefont {{Dodds-Eden}},
  \citenamefont {{Gerhard}}, \citenamefont {{Levin}}, \citenamefont
  {{Mascetti}}, \citenamefont {{Nayakshin}}, \citenamefont {{Perets}},
  \citenamefont {{Perrin}}, \citenamefont {{Pfuhl}}, \citenamefont {{Reid}},
  \citenamefont {{Rouan}}, \citenamefont {{Zilka}},\ and\ \citenamefont
  {{Sternberg}}}]{BartkoEtAl10}%
  \BibitemOpen
  \bibfield  {author} {\bibinfo {author} {\bibfnamefont {H.}~\bibnamefont
  {{Bartko}}}, \bibinfo {author} {\bibfnamefont {F.}~\bibnamefont {{Martins}}},
  \bibinfo {author} {\bibfnamefont {S.}~\bibnamefont {{Trippe}}}, \bibinfo
  {author} {\bibfnamefont {T.~K.}\ \bibnamefont {{Fritz}}}, \bibinfo {author}
  {\bibfnamefont {R.}~\bibnamefont {{Genzel}}}, \bibinfo {author}
  {\bibfnamefont {T.}~\bibnamefont {{Ott}}}, \bibinfo {author} {\bibfnamefont
  {F.}~\bibnamefont {{Eisenhauer}}}, \bibinfo {author} {\bibfnamefont
  {S.}~\bibnamefont {{Gillessen}}}, \bibinfo {author} {\bibfnamefont
  {T.}~\bibnamefont {{Paumard}}}, \bibinfo {author} {\bibfnamefont
  {T.}~\bibnamefont {{Alexander}}}, \bibinfo {author} {\bibfnamefont
  {K.}~\bibnamefont {{Dodds-Eden}}}, \bibinfo {author} {\bibfnamefont
  {O.}~\bibnamefont {{Gerhard}}}, \bibinfo {author} {\bibfnamefont
  {Y.}~\bibnamefont {{Levin}}}, \bibinfo {author} {\bibfnamefont
  {L.}~\bibnamefont {{Mascetti}}}, \bibinfo {author} {\bibfnamefont
  {S.}~\bibnamefont {{Nayakshin}}}, \bibinfo {author} {\bibfnamefont {H.~B.}\
  \bibnamefont {{Perets}}}, \bibinfo {author} {\bibfnamefont {G.}~\bibnamefont
  {{Perrin}}}, \bibinfo {author} {\bibfnamefont {O.}~\bibnamefont {{Pfuhl}}},
  \bibinfo {author} {\bibfnamefont {M.~J.}\ \bibnamefont {{Reid}}}, \bibinfo
  {author} {\bibfnamefont {D.}~\bibnamefont {{Rouan}}}, \bibinfo {author}
  {\bibfnamefont {M.}~\bibnamefont {{Zilka}}}, \ and\ \bibinfo {author}
  {\bibfnamefont {A.}~\bibnamefont {{Sternberg}}},\ }\href {\doibase
  10.1088/0004-637X/708/1/834} {\bibfield  {journal} {\bibinfo  {journal}
  {ApJ}\ }\textbf {\bibinfo {volume} {708}},\ \bibinfo {pages} {834} (\bibinfo
  {year} {2010})},\ \Eprint {http://arxiv.org/abs/0908.2177} {arXiv:0908.2177}
  \BibitemShut {NoStop}%
\bibitem [{\citenamefont {{Ballero}}\ \emph {et~al.}(2007)\citenamefont
  {{Ballero}}, \citenamefont {{Kroupa}},\ and\ \citenamefont
  {{Matteucci}}}]{BalleroEtAl2007}%
  \BibitemOpen
  \bibfield  {author} {\bibinfo {author} {\bibfnamefont {S.~K.}\ \bibnamefont
  {{Ballero}}}, \bibinfo {author} {\bibfnamefont {P.}~\bibnamefont {{Kroupa}}},
  \ and\ \bibinfo {author} {\bibfnamefont {F.}~\bibnamefont {{Matteucci}}},\
  }\href {\doibase 10.1051/0004-6361:20066786} {\bibfield  {journal} {\bibinfo
  {journal} {A\&A}\ }\textbf {\bibinfo {volume} {467}},\ \bibinfo {pages} {117}
  (\bibinfo {year} {2007})},\ \Eprint {http://arxiv.org/abs/astro-ph/0702047}
  {astro-ph/0702047} \BibitemShut {NoStop}%
\bibitem [{\citenamefont {{Sch{\"o}del}}\ \emph {et~al.}(2014)\citenamefont
  {{Sch{\"o}del}}, \citenamefont {{Feldmeier}}, \citenamefont {{Neumayer}},
  \citenamefont {{Meyer}},\ and\ \citenamefont {{Yelda}}}]{SchoedelEtAl2014}%
  \BibitemOpen
  \bibfield  {author} {\bibinfo {author} {\bibfnamefont {R.}~\bibnamefont
  {{Sch{\"o}del}}}, \bibinfo {author} {\bibfnamefont {A.}~\bibnamefont
  {{Feldmeier}}}, \bibinfo {author} {\bibfnamefont {N.}~\bibnamefont
  {{Neumayer}}}, \bibinfo {author} {\bibfnamefont {L.}~\bibnamefont {{Meyer}}},
  \ and\ \bibinfo {author} {\bibfnamefont {S.}~\bibnamefont {{Yelda}}},\ }\href
  {\doibase 10.1088/0264-9381/31/24/244007} {\bibfield  {journal} {\bibinfo
  {journal} {Classical and Quantum Gravity}\ }\textbf {\bibinfo {volume}
  {31}},\ \bibinfo {eid} {244007} (\bibinfo {year} {2014})},\ \Eprint
  {http://arxiv.org/abs/1411.4504} {arXiv:1411.4504} \BibitemShut {NoStop}%
\bibitem [{\citenamefont {{Bahcall}}\ and\ \citenamefont
  {{Wolf}}(1977)}]{BW77}%
  \BibitemOpen
  \bibfield  {author} {\bibinfo {author} {\bibfnamefont {J.~N.}\ \bibnamefont
  {{Bahcall}}}\ and\ \bibinfo {author} {\bibfnamefont {R.~A.}\ \bibnamefont
  {{Wolf}}},\ }\href@noop {} {\bibfield  {journal} {\bibinfo  {journal} {ApJ}\
  }\textbf {\bibinfo {volume} {216}},\ \bibinfo {pages} {883} (\bibinfo {year}
  {1977})}\BibitemShut {NoStop}%
\bibitem [{\citenamefont {{Finn}}\ and\ \citenamefont
  {{Thorne}}(2000)}]{FinnThorne2000}%
  \BibitemOpen
  \bibfield  {author} {\bibinfo {author} {\bibfnamefont {L.~S.}\ \bibnamefont
  {{Finn}}}\ and\ \bibinfo {author} {\bibfnamefont {K.~S.}\ \bibnamefont
  {{Thorne}}},\ }\href {\doibase 10.1103/PhysRevD.62.124021} {\bibfield
  {journal} {\bibinfo  {journal} {Phys. Rev. D}\ }\textbf {\bibinfo {volume}
  {62}},\ \bibinfo {eid} {124021} (\bibinfo {year} {2000})},\ \Eprint
  {http://arxiv.org/abs/gr-qc/0007074} {gr-qc/0007074} \BibitemShut {NoStop}%
\bibitem [{\citenamefont {{Peters}}(1964)}]{Peters64}%
  \BibitemOpen
  \bibfield  {author} {\bibinfo {author} {\bibfnamefont {P.~C.}\ \bibnamefont
  {{Peters}}},\ }\href@noop {} {\bibfield  {journal} {\bibinfo  {journal}
  {Physical Review}\ }\textbf {\bibinfo {volume} {136}},\ \bibinfo {pages}
  {1224} (\bibinfo {year} {1964})}\BibitemShut {NoStop}%
\bibitem [{\citenamefont
  {{Amaro-Seoane}}(2018{\natexlab{c}})}]{Amaro-SeoaneLRR}%
  \BibitemOpen
  \bibfield  {author} {\bibinfo {author} {\bibfnamefont {P.}~\bibnamefont
  {{Amaro-Seoane}}},\ }\href {\doibase 10.1007/s41114-018-0013-8} {\bibfield
  {journal} {\bibinfo  {journal} {Living Reviews in Relativity}\ }\textbf
  {\bibinfo {volume} {21}},\ \bibinfo {eid} {4} (\bibinfo {year}
  {2018}{\natexlab{c}})}\BibitemShut {NoStop}%
\bibitem [{\citenamefont {{Alexander}}\ and\ \citenamefont
  {{Livio}}(2001)}]{AL01}%
  \BibitemOpen
  \bibfield  {author} {\bibinfo {author} {\bibfnamefont {T.}~\bibnamefont
  {{Alexander}}}\ and\ \bibinfo {author} {\bibfnamefont {M.}~\bibnamefont
  {{Livio}}},\ }\href@noop {} {\bibfield  {journal} {\bibinfo  {journal} {ApJ
  Lett.}\ }\textbf {\bibinfo {volume} {560}},\ \bibinfo {pages} {143} (\bibinfo
  {year} {2001})}\BibitemShut {NoStop}%
\bibitem [{\citenamefont {{Spitzer}}(1987)}]{Spitzer87}%
  \BibitemOpen
  \bibfield  {author} {\bibinfo {author} {\bibfnamefont {L.}~\bibnamefont
  {{Spitzer}}},\ }\href@noop {} {\emph {\bibinfo {title} {{Dynamical evolution
  of globular clusters}}}}\ (\bibinfo  {publisher} {Princeton, NJ, Princeton
  University Press, 1987, 191 p.},\ \bibinfo {year} {1987})\BibitemShut
  {NoStop}%
\bibitem [{\citenamefont {{Freitag}}\ \emph {et~al.}(2006)\citenamefont
  {{Freitag}}, \citenamefont {{Amaro-Seoane}},\ and\ \citenamefont
  {{Kalogera}}}]{FAK06a}%
  \BibitemOpen
  \bibfield  {author} {\bibinfo {author} {\bibfnamefont {M.}~\bibnamefont
  {{Freitag}}}, \bibinfo {author} {\bibfnamefont {P.}~\bibnamefont
  {{Amaro-Seoane}}}, \ and\ \bibinfo {author} {\bibfnamefont {V.}~\bibnamefont
  {{Kalogera}}},\ }\href {\doibase 10.1086/506193} {\bibfield  {journal}
  {\bibinfo  {journal} {ApJ}\ }\textbf {\bibinfo {volume} {649}},\ \bibinfo
  {pages} {91} (\bibinfo {year} {2006})},\ \Eprint
  {http://arxiv.org/abs/arXiv:astro-ph/0603280} {arXiv:astro-ph/0603280}
  \BibitemShut {NoStop}%
\bibitem [{\citenamefont {{Bianchini}}\ \emph {et~al.}(2016)\citenamefont
  {{Bianchini}}, \citenamefont {{van de Ven}}, \citenamefont {{Norris}},
  \citenamefont {{Schinnerer}},\ and\ \citenamefont
  {{Varri}}}]{BianchiniEtAl2016}%
  \BibitemOpen
  \bibfield  {author} {\bibinfo {author} {\bibfnamefont {P.}~\bibnamefont
  {{Bianchini}}}, \bibinfo {author} {\bibfnamefont {G.}~\bibnamefont {{van de
  Ven}}}, \bibinfo {author} {\bibfnamefont {M.~A.}\ \bibnamefont {{Norris}}},
  \bibinfo {author} {\bibfnamefont {E.}~\bibnamefont {{Schinnerer}}}, \ and\
  \bibinfo {author} {\bibfnamefont {A.~L.}\ \bibnamefont {{Varri}}},\ }\href
  {\doibase 10.1093/mnras/stw552} {\bibfield  {journal} {\bibinfo  {journal}
  {MNRAS}\ }\textbf {\bibinfo {volume} {458}},\ \bibinfo {pages} {3644}
  (\bibinfo {year} {2016})},\ \Eprint {http://arxiv.org/abs/1603.00878}
  {arXiv:1603.00878} \BibitemShut {NoStop}%
\bibitem [{\citenamefont {{Amaro-Seoane}}\ \emph {et~al.}(2013)\citenamefont
  {{Amaro-Seoane}}, \citenamefont {{Sopuerta}},\ and\ \citenamefont
  {{Freitag}}}]{Amaro-SeoaneSopuertaFreitag2013}%
  \BibitemOpen
  \bibfield  {author} {\bibinfo {author} {\bibfnamefont {P.}~\bibnamefont
  {{Amaro-Seoane}}}, \bibinfo {author} {\bibfnamefont {C.~F.}\ \bibnamefont
  {{Sopuerta}}}, \ and\ \bibinfo {author} {\bibfnamefont {M.~D.}\ \bibnamefont
  {{Freitag}}},\ }\href {\doibase 10.1093/mnras/sts572} {\bibfield  {journal}
  {\bibinfo  {journal} {MNRAS}\ }\textbf {\bibinfo {volume} {429}},\ \bibinfo
  {pages} {3155} (\bibinfo {year} {2013})},\ \Eprint
  {http://arxiv.org/abs/1205.4713} {arXiv:1205.4713 [astro-ph.CO]} \BibitemShut
  {NoStop}%
\bibitem [{\citenamefont {{Peters}}\ and\ \citenamefont
  {{Mathews}}(1963)}]{PM63}%
  \BibitemOpen
  \bibfield  {author} {\bibinfo {author} {\bibfnamefont {P.~C.}\ \bibnamefont
  {{Peters}}}\ and\ \bibinfo {author} {\bibfnamefont {J.}~\bibnamefont
  {{Mathews}}},\ }\href@noop {} {\bibfield  {journal} {\bibinfo  {journal}
  {Physical Review}\ }\textbf {\bibinfo {volume} {131}},\ \bibinfo {pages}
  {435} (\bibinfo {year} {1963})}\BibitemShut {NoStop}%
\bibitem [{\citenamefont {Alexander}\ and\ \citenamefont
  {Hopman}(2009)}]{AlexanderHopman09}%
  \BibitemOpen
  \bibfield  {author} {\bibinfo {author} {\bibfnamefont {T.}~\bibnamefont
  {Alexander}}\ and\ \bibinfo {author} {\bibfnamefont {C.}~\bibnamefont
  {Hopman}},\ }\href {\doibase 10.1088/0004-637X/697/2/1861} {\bibfield
  {journal} {\bibinfo  {journal} {ApJ}\ }\textbf {\bibinfo {volume} {697}},\
  \bibinfo {pages} {1861} (\bibinfo {year} {2009})}\BibitemShut {NoStop}%
\bibitem [{\citenamefont {{Preto}}\ and\ \citenamefont
  {{Amaro-Seoane}}(2010)}]{PretoAmaroSeoane10}%
  \BibitemOpen
  \bibfield  {author} {\bibinfo {author} {\bibfnamefont {M.}~\bibnamefont
  {{Preto}}}\ and\ \bibinfo {author} {\bibfnamefont {P.}~\bibnamefont
  {{Amaro-Seoane}}},\ }\href {\doibase 10.1088/2041-8205/708/1/L42} {\bibfield
  {journal} {\bibinfo  {journal} {ApJ Lett.}\ }\textbf {\bibinfo {volume}
  {708}},\ \bibinfo {pages} {L42} (\bibinfo {year} {2010})},\ \Eprint
  {http://arxiv.org/abs/0910.3206} {arXiv:0910.3206} \BibitemShut {NoStop}%
\bibitem [{\citenamefont {{Amaro-Seoane}}\ and\ \citenamefont
  {{Preto}}(2011)}]{Amaro-SeoanePreto11}%
  \BibitemOpen
  \bibfield  {author} {\bibinfo {author} {\bibfnamefont {P.}~\bibnamefont
  {{Amaro-Seoane}}}\ and\ \bibinfo {author} {\bibfnamefont {M.}~\bibnamefont
  {{Preto}}},\ }\href {\doibase 10.1088/0264-9381/28/9/094017} {\bibfield
  {journal} {\bibinfo  {journal} {Classical and Quantum Gravity}\ }\textbf
  {\bibinfo {volume} {28}},\ \bibinfo {pages} {094017} (\bibinfo {year}
  {2011})},\ \Eprint {http://arxiv.org/abs/1010.5781} {arXiv:1010.5781
  [astro-ph.CO]} \BibitemShut {NoStop}%
\bibitem [{\citenamefont {{Emami}}\ and\ \citenamefont
  {{Loeb}}(2019)}]{EmamiLoeb2019}%
  \BibitemOpen
  \bibfield  {author} {\bibinfo {author} {\bibfnamefont {R.}~\bibnamefont
  {{Emami}}}\ and\ \bibinfo {author} {\bibfnamefont {A.}~\bibnamefont
  {{Loeb}}},\ }\href@noop {} {\bibfield  {journal} {\bibinfo  {journal} {arXiv
  e-prints}\ ,\ \bibinfo {eid} {arXiv:1903.02579}} (\bibinfo {year} {2019})},\
  \Eprint {http://arxiv.org/abs/1903.02579} {arXiv:1903.02579 [astro-ph.HE]}
  \BibitemShut {NoStop}%
\end{thebibliography}
\end{document}